\documentclass[acmsmall]{acmart}
\usepackage{graphicx}
\usepackage{amsmath}
\usepackage{booktabs}
\usepackage{algorithmic}
\usepackage{cite}
\usepackage{multirow}
\usepackage{url}
\usepackage{bm}
\usepackage{amsmath,amsfonts}
\usepackage{graphicx}
\usepackage{textcomp}
\usepackage{xcolor}
\usepackage{amsthm}
\usepackage{subfigure}
\usepackage{stfloats}
\usepackage{caption}
\usepackage{paralist}
\usepackage{chngcntr}
\usepackage{kotex}
\AtBeginDocument{%
  \providecommand\BibTeX{{%
    \normalfont B\kern-0.5em{\scshape i\kern-0.25em b}\kern-0.8em\TeX}}}

\begin{document}

%%
%% The "title" command has an optional parameter,
%% allowing the author to define a "short title" to be used in page headers.
\title{Human Mobility during COVID-19 in the Context of Mild\\Social Distancing: Implications for Technological Interventions}

%%%%%%%%%%% Alternative title: How did COVID-19 influence human mobility?
%%% Micro-level Human Mobility during COVID-19: The Context of Mild Social Distancing

%%
%% The "author" command and its associated commands are used to define
%% the authors and their affiliations.
%% Of note is the shared affiliation of the first two authors, and the
%% "authornote" and "authornotemark" commands
%% used to denote shared contribution to the research.
\author{Myeong Lee}
\email{mlee89@gmu.edu}
\authornote{Both authors contributed equally to this research.}
\affiliation{%
  \institution{Dept. of Information Sciences and Technology, George Mason University}
  \streetaddress{4400 University Dr.}
  \city{Fairfax}
  \state{VA}
  \postcode{22030}
  \country{United States}
}

\author{Seongkyu Lee}
\authornotemark[1]
\affiliation{%
  \institution{Dept. of Computer Science, Yonsei University}
  \streetaddress{50 Yonsei-ro Seodaemun-gu}
  \city{Seoul}
  \country{South Korea}}
\email{seongkyu.lee@yonsei.ac.kr}

\author{Seonghoon Kim}
\affiliation{%
  \institution{Motov, Inc.}
  \streetaddress{119 Songdo-Munhwa-ro, Yeonsu-gu}
  \city{Incheon}
  \country{South Korea}}
\email{seonghoon.kim@motov.co.kr}

\author{Noseong Park}
\authornote{Corresponding author.}
\affiliation{%
  \institution{Dept. of Artificial Intelligence, Yonsei University}
  \streetaddress{50 Yonsei-ro Seodaemun-gu}
  \city{Seoul}
  \country{South Korea}}
\email{noseong@yonsei.ac.kr}

% \author{Aparna Patel}
% \affiliation{%
%  \institution{Rajiv Gandhi University}
%  \streetaddress{Rono-Hills}
%  \city{Doimukh}
%  \state{Arunachal Pradesh}
%  \country{India}}

% \author{Huifen Chan}
% \affiliation{%
%   \institution{Tsinghua University}
%   \streetaddress{30 Shuangqing Rd}
%   \city{Haidian Qu}
%   \state{Beijing Shi}
%   \country{China}}

% \author{Charles Palmer}
% \affiliation{%
%   \institution{Palmer Research Laboratories}
%   \streetaddress{8600 Datapoint Drive}
%   \city{San Antonio}
%   \state{Texas}
%   \postcode{78229}}
% \email{cpalmer@prl.com}

% \author{John Smith}
% \affiliation{\institution{The Th{\o}rv{\"a}ld Group}}
% \email{jsmith@affiliation.org}

% \author{Julius P. Kumquat}
% \affiliation{\institution{The Kumquat Consortium}}
% \email{jpkumquat@consortium.net}

%%
%% By default, the full list of authors will be used in the page
%% headers. Often, this list is too long, and will overlap
%% other information printed in the page headers. This command allows
%% the author to define a more concise list
%% of authors' names for this purpose.
% \renewcommand{\shortauthors}{Trovato and Tobin, et al.}

%%
%% The abstract is a short summary of the work to be presented in the
%% article.
\begin{abstract}
The COVID-19 pandemic has brought both tangible and intangible damage to our society. Many researchers studied about its societal impacts in the countries that had implemented strong social distancing measures such as stay-at-home orders. Among them, human mobility has been studied extensively due to its importance in flattening the curve. However, mobility has not been actively studied in the context of mild social distancing. Insufficient understanding of human mobility in diverse contexts might provide limited implications for any technological interventions to alleviate the situation. To this end, we collected a dataset consisting of more than 1M daily smart device users in the third-largest city of South Korea, which has implemented mild social distancing policies. We analyze how COVID-19 shaped human mobility in the city from geographical, socio-economic, and socio-political perspectives. We also examine mobility changes for points of interest and special occasions such as transportation stations and the case of legislative elections. We identify a typology of populations through these analyses as a means to provide design implications for technological interventions. This paper contributes to social sciences through in-depth analyses of human mobility and to the CSCW community with new design challenges and potential implications. 
\end{abstract}

%%
%% The code below is generated by the tool at http://dl.acm.org/ccs.cfm.
%% Please copy and paste the code instead of the example below.
%%
% \begin{CCSXML}
% <ccs2012>
%  <concept>
%   <concept_id>10010520.10010553.10010562</concept_id>
%   <concept_desc>Computer systems organization~Embedded systems</concept_desc>
%   <concept_significance>500</concept_significance>
%  </concept>
%  <concept>
%   <concept_id>10010520.10010575.10010755</concept_id>
%   <concept_desc>Computer systems organization~Redundancy</concept_desc>
%   <concept_significance>300</concept_significance>
%  </concept>
%  <concept>
%   <concept_id>10010520.10010553.10010554</concept_id>
%   <concept_desc>Computer systems organization~Robotics</concept_desc>
%   <concept_significance>100</concept_significance>
%  </concept>
%  <concept>
%   <concept_id>10003033.10003083.10003095</concept_id>
%   <concept_desc>Networks~Network reliability</concept_desc>
%   <concept_significance>100</concept_significance>
%  </concept>
% </ccs2012>
% \end{CCSXML}

% \ccsdesc[500]{Computer systems organization~Embedded systems}
% \ccsdesc[300]{Computer systems organization~Redundancy}
% \ccsdesc{Computer systems organization~Robotics}
% \ccsdesc[100]{Networks~Network reliability}

%%
%% Keywords. The author(s) should pick words that accurately describe
%% the work being presented. Separate the keywords with commas.
\keywords{COVID-19, Human Mobility, Design Implications}
\maketitle

%%
%% This command processes the author and affiliation and title
%% information and builds the first part of the formatted document.

\section{Introduction}\label{sec:intro}

The recent COVID-19 pandemic (SARS-CoV-2) has caused radical social changes world-wide, driven by both citizens' prevention efforts and health authorities' policies and recommendations for flattening the curve. 
As part of such efforts, technological interventions such as mobility-tracking and self-diagnosis apps have been developed to help mitigate the risks (e.g., \citep{mayor2020covid,cho2020contact}).
While the context could be different from pandemics, crisis management strategies and technological interventions have been widely studied in a similar manner in the computer-supported cooperated work (CSCW), human-computer interaction (HCI), and a broader field of crisis informatics (e.g., \citep{perry2001dealing}). 
For example, scholars have studied information and communication technologies (ICTs) for disasters and emergencies, which include, but are not limited to, effective communications tools \citep{frommberger2013mobile4d}, people's re-appropriation of existing ICTs \citep{reuter2017social}, and crowdsourcing technologies for curating crisis-related data \citep{yang2014providing}.

To make these kinds of technological interventions effective for pandemics, understanding human mobility needs to be preceded, because the spread of the virus is closely related to human mobility \citep{kraemer2020effect}.
Due to this reason, social distancing has been one of the major guidelines for citizens in many countries, among other measures during the COVID-19 outbreak. 
Human mobility is a strong predictor of the epidemic diffusion in its early stage, but it is also a direct indicator for measuring people's social distancing practices \citep{kraemer2020effect}. 
By monitoring mobility in local communities, health authorities can benefit by having data-driven evidence to adjust contingency plans and prepare strategies for future pandemics dynamically. 

From a technological intervention perspective, the context of COVID-19 poses new challenges for CSCW scholars in understanding the new norms and mobility with respect to people's work and life. 
While outdoor activities are not recommended during the pandemic, it is still necessary for many people to work either from their homes or workplaces, to go for grocery shopping, and to do exercises for their health. 
Because the forms of these essential activities vary significantly by region, available resources, and even people's demographic characteristics \citep{coven2020disparities}, people's behavioral patterns shaped by social determinants might be different by social, geographical, and cultural groups. 
This is why understanding human mobility during pandemics is a basis for implementing health policies and augmenting technological interventions to flatten the curve as well as to build the community resilience capacity.
Without understanding the dynamics of human mobility that are often contingent on people's socio-economic and socio-cultural characteristics, a technological intervention that is effective for one group could be a source of inequality for another. 

In light of this implication, there have been many studies that focus on mobility changes during the COVID-19 outbreak.
On the one hand, research has examined national-, state-, or county-level mobility changes, which tend to reflect people's mid- to long-term mobility during the pandemic (e.g., international students and employees go back to their home countries for the pandemic period) \citep{klein2020assessing,google_mobilty}. 
On the other hand, with the support of IT companies and non-profit organizations that have curated mobility data of their customers, many others studied mobility changes at the granular geospatial levels, such as point of interest (POIs) \citep{mckenziea2020country}. 
These studies and statistics show where hot or dead spots are located during the pandemic and how people change their behaviors in a high geographical resolution, such as at a census block or a location level. 
They can also capture people's daily mobility patterns under strong social-distancing measures such as travel restriction \citep{klein2020assessing,zhang2020interactive}. 
However, these studies' implications could be limited when it comes to a context where social and economic activities are weakly restricted. 

To understand the role of socio-political context during the COVID-19 outbreak, we study the human mobility change between December in 2019 and May in 2020 in a metropolitan city in South Korea, where social distancing was encouraged but not implemented in the form of strong travel restrictions or stay-at-home orders. 
This mild social distancing was possible partially by virtue of the health authority's aggressive management in the early stage of the outbreak \citep{Beaubien_2020}. 
From a scientific point of view, the context of mild social distancing in South Korea provides opportunities to understand: (1) how human mobility manifests based on people's prevention behaviors and perceptions, rather than by the government's policies and administrative enforcement, (2) how social determinants of health outcomes such as demographic and socio-economic features shape people's mobility, (3) how mobility changes when the infection rate is decreasing (i.e., during a releasing period), and (4) what the mobility changes during special occasions such as the national election look like during the pandemic. 

In addition to these scientific contributions, this paper also provides design implications for pandemic-related technological systems. 
Beyond crisis management, the CSCW community has a long tradition of studying technology-assisted collaborative work for mobile workers \citep{perry2001dealing} and crowdsourcing strategies for solving community problems through spatio-temporal data \citep{liu2014crisis,priedhorsky2008computational}. 
Understanding human mobility during the pandemic is useful not only for the design of crisis management systems but also by suggesting a typology of potential users who respond differently to the pandemic. 
It also provides implications for technology-assisted collaborative work strategies such as collaborative problem-solving in distant work settings.
The mobility patterns of workers and their new norms in workplaces and homes might raise new challenges for collaborative systems, which might require a re-design of the existing systems. 
Unpacking these design challenges based on people's mobility patterns will help develop mobile working and crowdsourcing strategies during the pandemic. 
We expect that exploring this design space will facilitate discussions about the collaborative system design for the new norms after the pandemic by introducing a new set of behavioral characteristics that manifest in diverse socio-political contexts. 

To this end, we make use of the 5-month pedestrian location data in the city of Incheon where the largest international airport is located.\footnote{The population in the city of Incheon is around 3M. Due to its location next to Seoul, there are about 26M residents in a 2-hour distance from the city, characterizing it as a metropolitan area. It also has the fourth biggest number of COVID-19 confirmed cases in South Korea (see \url{http://ncov.mohw.go.kr/en}).} 
The data collection period is from December 18th in 2019 through April 30th in 2020, which covers the periods of pre-pandemic, the major COVID-19 outbreak, and relief.\footnote{Although the pandemic has not ended and its risk remains as of May 2020, the paper calls the period after the Mid-March through April as ``relieved'' based on news articles' characterization of the period. See \ref{sec:events}.} 
This dataset is a by-product of the location-based advertisement recommendation systems and was made available based on a partnership between the authors' institutions and a start-up company in South Korea. 

We aim to understand human mobility during COVID-19 in the context of mild social distancing through this data analysis. 
Based on this analysis, we identify a typology of populations from a mobility perspective during the pandemic: \emph{crowd-avoiding outdoor workers, old workers, working voters, flexible office workers, and leisure-time seekers}.
This typology provides design implications for technological interventions during pandemics.

% By analyzing this data, we show that (1) many areas show a significant mobility decrease/increase during pandemic, (2) urban areas with higher poverty show a larger decrease in mobility compared to other regions, and (3) pandemic-related events such as a huge outbreak in other city and governments' recommendations can change mobility. 

\section{Human Mobility during the Pandemic}

We first review scientific research about human mobility during the pandemic in diverse contexts. 
Mobility research can be classified into macro-, meso- and microscopic analyses depending on the geographical unit of analysis and the mobility data aggregation level \citep{klein2020assessing}.
The macroscopic-level analysis includes studies that focus on international, inter-state, or inter-county mobility patterns, which might help understand mid- to long-term migration behaviors rather than daily mobility patterns. 
The mesoscopic level covers regions smaller than a county or metropolitan area, such as within a Combined Statistical Area (CSA) of the United States Census. 
Such analyses focus on small-region mobility, such as cross-census tract dynamics. 
Finally, microscopic-level analyses focus on place-level or location-based analysis where the resolution of the geographical unit is precise geo-coordinates. 
Along with this classification, we review pandemic-related mobility research in several socio-political contexts related to the pandemic.
Subsequently, we review known risk factors and social determinants of health outcomes that can give rise to human mobility.

\subsection{Mobility under Strong Measures}

Many studies about human mobility during the COVID-19 outbreak have focused on regions or countries that implemented a strong social distancing measure such as a stay-at-home order for a majority population.
The analysis level in this context ranges from macroscopic- to microscopic-level analyses. 
Many of them made use of mobility data provided by IT companies or non-profit organizations that was made available to the public or on-demand as part of efforts to overcome the pandemic. 
These studies examined the effects of strong social distancing measures in different governance units. 
For example, a macroscopic-level study using the mobility data from Baidu showed that the inter-city and inter-province mobility in China was a proxy of the spread of the virus in the early stage of the pandemic \citep{kraemer2020effect}. 
This study concluded that the correlation between inter-city mobility and the spread of the virus weakened significantly after the country had taken a strong measure of locking down the affected cities. 

While the degree of enforcement was different from that in China, studies in the contexts of North America and Europe provide similar implications for the impact of social distancing measures. 
Del Fava et al. showed decreasing patterns in social contacts by directly surveying people during COVID-19 and found that the reduction in social contacts after announcing the social distancing guideline was smaller in North America and European countries, except for Italy than China \citep{del2020differential}. 
Similarly, by making use of the Cuebiq data, Klein at al. found that, after the stay-at-home order in multiple states in the United States, inter-CSA travels decreased by 60\% to 80\% (macroscopic) and inter-census tract mobility decreased by 50\% (mesoscopic). Individual-level mobility measured using the radius of gyration decreased by 40\% to 60\% during weekdays (microscopic) \citep{klein2020assessing}.
Using the same dataset but focusing on the New York metropolitan area, Bakker et al. reported similar patterns at the place level, showing that individuals' average travel distance per day decreased by 70\% during weekends, the social contact rate decreased by 93\%, and the place visiting rate decreased by 60\% after the national emergency declaration in the area \citep{bakker2020effect}.
Using the same mobility data, similar findings were found in Italy, in which a strong social distancing measure was implemented \citep{pepe2020covid}.

Studies with other datasets confirmed the same findings as that of the studies using the Cuebiq data. 
For example, Descartes Lab. curated human mobility data from smartphone users' GPS logs \citep{warren2020mobility}.
This study reported sharp mobility drops in many states in the United States after strong measures and explained some contextualized patterns by regions (macroscopic). 
Combining this data with the SafeGraph dataset, another study also developed a GIS-based visualization system for monitoring county-level human mobility, showing the change of people's dwelling time by county over time \citep{gao2020mapping}.
A study that used the Unacast data reported that, at the county-level, a reduction in mobility was associated with a rise in the local infection rate \citep{engle2020staying}, which is consistent with the findings of other studies \citep{kraemer2020effect,gao2020mobile}.

Overall, these studies show the effects of strong social distancing measures on human mobility in various locations, regions, and countries.
Although the administrative management of health authorities do affect human mobility, people's voluntary efforts and prevention behaviors also give rise to their mobility changes \citep{warren2020mobility}.
This phenomenon can be observed better when social distancing measures are weakly implemented because government policies may have had relatively small effects on human mobility. 

\subsection{Mobility under Mild Measures}

The number of countries that took a mild social distancing measure is smaller than those that took a strong measure. 
Countries that were studied in the context of mild social distancing are Sweden and South Korea \citep{dahlberg2020effects,park2020changes}.
Although both countries took similar measures, their backgrounds and motivations are different. 
According to a news article, the ``Swedish model,'' which keeps schools and other facilities open while encouraging social distancing, is intended to maintain the national healthcare system's stability by sharing the burden of the healthcare workers in taking care of their children \citep{tharoor_2020}.
This article reported that the Swedish model was implemented based on people's strong trust towards the health authorities.  
Meanwhile, in South Korea, a mild social distancing was implemented, aided by their testings to trace all individuals who contacted confirmed patients \citep{Beaubien_2020}.
While taking a strong measure of self-quarantine and converted schools online, Korean governments have allowed free movement for most of the population. 
Because of these differences between the two countries, their meso- or microscopic mobility patterns might be different.  

To our knowledge, only several studies have analyzed mobility during COVID-19 in the context of mild social distancing. 
% A lack of studies in this context is one of our motivations to examine mesoscopic human mobility in a metropolitan city in South Korea. 
A study that used call detail records (CDR) acquired by a telecom company in Sweden reported that there was a 64\% decrease in mobility in residential areas (mesoscopic), a 33\% decrease in commercial areas during daytime (mesoscopic), and a 38\% decrease in the maximum travel distance per day (microscopic) in the greater Stockholm region \citep{dahlberg2020effects}. 
Another study using the Seoul Metropolitan Subway data in South Korea reported that daily passengers decreased by 40.6\% overall (mesoscopic) \citep{park2020changes}.

These studies provide a similar mobility change implication to that in the countries with strong measures: (1) the overall mobility in metropolitan areas decreases significantly; (2) the mobility in commercial areas decreases relatively less compared to residential areas, maybe due to the continued operations of businesses; and (3) demographic and socio-economic features minimally give rise to the variation of mobility change. 
Although these studies analyzed mobility changes during COVID-19 with respect to demographic and socio-economic status at the micro- and mesoscopic levels, their analyses are limited due to their short analysis period, implicit classification of land-use, and low temporal resolution.
The Sweden case focused on daytime mobility during an early stage of the COVID-19 development, and the South Korean case focused on only subway transit logs for a small number of subway stations. Due to these limitations, more systematic and contextualized analyses are necessary to understand how mobility patterns change and manifest in the context of mild social distancing. 

To provide better implications for mobility changes in this context, we aim to conduct analyses by land-use types, socio-economic factors, and demographic features at the mesoscopic region level. 
As the first step toward the goal, we ask the following Research Question 1: 

\begin{itemize}
    \item \emph{RQ1: How did socio-political events about COVID-19 affect human mobility in South Korea?}\footnote{``Socio-political events'' mean any articles, governments' policies, and other events about COVID-19 that attracted attention. See Section~\ref{sec:events}.}
\end{itemize}

\subsection{Social Determinants of Health Outcomes and Mobility}\label{sec:determinants}

In addition to socio-political events, social determinants of health outcomes, such as demographic features, could be related to a risk of COVID-19, which in turn might affect people's preventive behaviors.
While chronic diseases and other health conditions are considered critical factors, this paper focuses on social determinants of health outcomes for our focus on community dynamics.
In public health, it is known that health outcomes such as mortality rate are heavily influenced by people's socio-economic status (e.g., income, education, and employment status), demographic features (e.g., age), and social capital (e.g., the strength of social support networks) \citep{havranek2015social}. 
Because these factors are closely related to or confounded with individuals' access to healthcare services, health-related behaviors, and nutrition levels, mitigating the potential inequality in local communities is one of the main goals on which many medical and public health experts have focused. 
During COVID-19, while it is still a developing phenomenon as of May 2020, individuals with high socio-economic deprivation were victimized most in the United States \citep{thebault_2020}.

Countries with mild distancing measures might present similar patterns to the United States when it comes to social determinants of health outcomes, even though their socio-economic and socio-cultural dynamics are different.
Of course, the healthcare systems in South Korea and Sweden are universal and government-funded, so the effect sizes of the social determinants could be different from that in the United States.
However, human mobility can still vary depending on socio-economic status in these countries because of the nature of different professions and occupations across socio-economic groups \citep{coven2020disparities}. 
Because it is still unclear whether socio-economic status shaped the mobility patterns of individuals under mild social distancing measures, we ask Research Question 2 as follows:

\begin{itemize}
    \item \emph{RQ2: How did human mobility in South Korea vary depending on socio-economic status?}
\end{itemize}

Demographic features are also known as important risk factors for COVID-19. 
Studies found that age and sex give rise to the degree of risk (mortality rate) \citep{caramelo2020estimation, jordan2020covid}.
Men were reported as being more vulnerable to COVID-19 than women and women tend to pursue safety more actively than men from a behavioral perspective \citep{perrotta2020behaviors}. 
This shows that gender-driven risk-taking behavior may mediate the effects of the reported medical risk on human mobility (i.e., the social construction of the reported risk).

Age also matters. 
The medical risk from COVID-19 is higher among older people; thus, the perceived risk of COVID-19 may be higher among them as well.
A study showed that the social distancing rate during the COVID-19 outbreak was higher among older people than young people in eight European countries \citep{del2020differential}.
Because the demographic factors correlated with the medical risk may shape people's perception of COVID-19 differently, human mobility might vary depending on the demographic characteristics. 
Overall, we expect that both the medical risk and the social-construction of gender-based prevention behaviors give rise to human mobility. 
Previous studies on sex as a biological factor \citep{jordan2020covid} and gender-based behavioral discrepancies \citep{perrotta2020behaviors} suggest two potentially-conflicting inferences for the effects of sex on mobility during COVID-19. 
Because the reported risk from COVID-19 is higher among men, we can hypothesize that the drop rate of men's mobility might be higher than that of women. 
Conversely, it is also reasonable to infer the other way around due to the behavioral patterns.
Therefore, we ask Research Question 3 as follows:

\begin{itemize}
    \item \emph{RQ3: How are demographic characteristics related to human mobility during COVID-19?}
\end{itemize}

\subsection{Points of Interest (POIs) and Mobility}

% Finally, we examine whether regions with important POIs such as international airport, transit stations, universities, hospitals, and so forth show any abnormal patterns during the pandemic. 
Several recent studies used the classification of places as a proxy for understanding the motivation of mobility changes at the microscopic level during COVID-19 (e.g., whether each individual's movement was for an essential or non-essential purpose) \citep{mckenziea2020country,klein2020assessing}. 
While partially providing similar implications, we study unique POIs that are related to, but are not limited to, the international travel through a hub airport, the domestic travel through major bus stations, and the national election day. 
There is the biggest international airport in South Korea in the City of Incheon. For example, the mobility change in the airport area can imply the volumes of international travel from and to South Korea.

Another major event during the pandemic was the national elections for congressmen and congresswomen, which were held on April 15th, 2020 \citep{olsen_2020}.
This was the first national elections in the world under COVID-19, which let the South Korean governments take new approaches. 
For example, the government employees made sure that voters keep social distancing while waiting on line, wear masks, and sanitize their hands throughout the voting process. 
Because of this new measure, human mobility on the election day would present a unique mobility pattern compared to other days. 

Because some of these POI-based analyses may show unique mobility patterns that can be observed only under the mild social-distancing measure, POI-based analyses might provide a nuanced understanding of the mobility changes during COVID-19.
In this regard, we ask Research Question 4 as follows:
\begin{itemize}
    \item \emph{RQ4: How did human mobility change during COVID-19 in the regions with important POIs and on the national election day?}
\end{itemize}

\section{Data Curation} 
%%% Please feel free to edit Seonghoon's part. You don't need any permission and any prior approval.
%Taxi's location data is captured every second and demographic data is curated every minute in an aggregated manner to protect user privacy. 

\begin{figure}[t]
	\centering
	\includegraphics[width=0.75\textwidth]{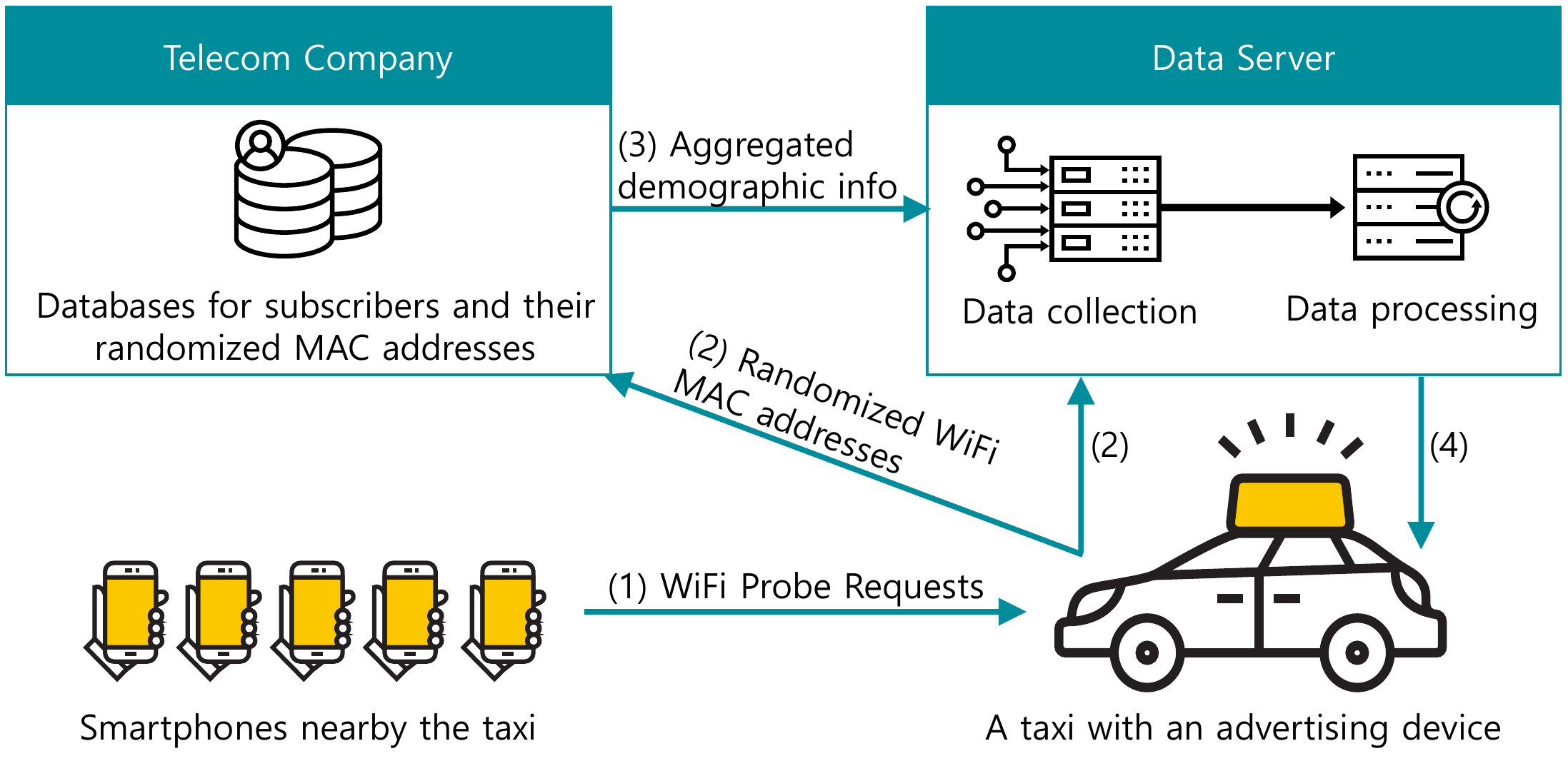}
	\caption{The data curation workflow. Smartphones send a probing message periodically to find nearby WiFi access points. The transmission delay of the probing message is under 50 microseconds in general, and the signal traverses through the air in the speed of light~\citep{10.1007/s11276-009-0192-z}. The signal decoding time in the collecting device is also under a few microseconds. Therefore, a probing message is received instantaneously, and data is curated in real-time, even at the taxi velocity of 100km/h.}
	\label{fig:data_curation}
\end{figure}

Data is curated by a location-based advertisement start-up in South Korea. 
This company works with one of the largest telecom companies in South Korea to detect real-time demographic information around a taxi.
Taxis use this demographic information to provide targeted advertisements that are streamed through the rooftop displays. 
% What follows describes how data is collected, curated, and used in detail.
Overall, there are more than 200 taxis deployed in the city of Incheon, the third-largest city in South Korea. 
Incheon International Airport, located in this city, carries 77\% of South Korea's international flights. 
Each taxi has a WiFi intelligent access point (IAP), a Long-term Evolution (LTE) cellular modem, a GPS.
% a variety of sensors, including accelerometer, gyroscope, thermometer, humidity sensor, luminance sensor, and passenger boarding sensor. 
IAP is a customized version of the general WiFi access point (AP) that captures MAC addresses of WiFi devices such as smartphones in real-time. 
To protect user privacy during the data acquisition process, the demographic information is collected in an aggregated manner, and MAC addresses are randomized. 
% We use the term ``MAC address'' to denote ``randomized MAC address'' throughout this paper.

The data curation process is as follows. The numbers for the following steps correspond to those in the workflow in Fig.~\ref{fig:data_curation}. 
(1) Individual smartphones keep sending WiFi probe requests to find a new AP. 
Because the IAP on a taxi keeps scanning these WiFi probe request messages, it can extract an MAC address from a request message. 
(2) The taxi sends the extracted MAC address to both the telecom company and the data server of the start-up company. 
(3) The telecom company looks up its subscriber database, as MAC addresses are received, and aggregates the demographic information (i.e., age and gender) every minute. 
Then, they send the aggregated demographic data to the start-up's server every minute. 
As a result, the returned demographic data does not include any original MAC addresses. 
The start-up company ensures privacy by making it impossible to join taxis' sensory data and the returned demographic data. 
Finally, (4) the returned demographic data is used for targeted advertisements that stream through the rooftop displays in real-time. 

The number of captured MAC addresses and their corresponding demographic information aggregated every minute are the key to understanding mobility in our analysis. 
\emph{Volume per minute} (VPM), denoted $p_{i,t}$ for taxi $i$ at time point $t$, is the unit of human mobility (see our notations in Table~\ref{tbl:notation}). 
% The data collection mechanism works well, even when a taxi moves fast. 
The quality of data collection was tested at up to 100km/h without failure, which ensures the data consistency and stability (see the description in Fig.~\ref{fig:data_curation}).

% (4) Together with MAC addresses, the advertising device sends to the data servers a sensor stream message collected every second. Each sensor stream message contains taxi location, speed, accelerometers, gyroscope, passenger boarding information (on/off), and system usage data. On receiving the messages, the data servers store them in database and run several batch processes to analyze and measure ad exposure efficiency by aggregating demographic information and sensor stream data every minute. (5) Also, they are used to better target audiences nearby taxis by adjusting ad exposure schedule and deploying ad creatives to each taxi.

%need something more???

%%%%%%%%%%%%%%%%%%% Seongkyu Lee
\section{Data Processing}
In this section, technical details are presented to show how key variables are generated and how the spatial and temporal units of analysis are determined. This process includes (1) the Voronoi decomposition~\citep{a71ff2df9f3246469efe8638d750498d} of the city of Incheon for determining the geospatial unit of analysis, (2) the temporal aggregation of the collected demographic data for determining the temporal unit of analysis, (3) the land-use classification in the city of Incheon, (4) the socio-political segmentation of the data collection period, denoted $\mathcal{T}$, and (5) the geographical imputation of socio-economic status at the Voronoi cells. We present key notations in Table~\ref{tbl:notation}.

\begin{table}[]
\centering
\footnotesize
\caption{The notation table}\label{tbl:notation}
\resizebox{\textwidth}{!}{\begin{tabular}{|c|c|}
\hline
Symbol  & Meaning \\ \hline
$i, t$ & Taxi $i$, time point $t$ in second (e.g., $t$ can be 23:59:59, 2020-03-13) \\ \hline
$D$ & A set of discrete time points\\ \hline
$\mathcal{T}$ & The entire time period in our dataset, i.e., from 00:00:00, 2019-12-18 to 23:59:59, 2020-04-30 \\ \hline
$t_f, t_m, t_g$ & The first outbreak time $t_f$, the first massive infection time $t_m$, and the golden-cross time $t_g$ \\ \hline
$t_0, t_l$ & The first and last time point of $\mathcal{T}$ in our dataset\\ \hline
$h$ & An hour band in a day. $h$ can be one of $\{[0,1), [1,2), \dots, [23,0)\}$.\\ \hline
$c, C$ & Voronoi cell $c$ and a cluster $C$ of Voronoi cells \\ \hline
$s_{i,t}$ & The <Longitude, Latitude> information of taxi $i$ at time point $t$ \\ \hline
$p_{i,t}$ & \begin{tabular}[c]{@{}c@{}} The volume of floating population detected by taxi $i$ for the past one minute from $t$, i.e., $[t - 60, t]$. \\We call this as \emph{volume per minute} (VPM). This is taught by the telecom company.\end{tabular}\\ \hline
$p_{c,D}$ & \begin{tabular}[c]{@{}c@{}} The average volume of per-minute floating population in cell $c$ at time points in $D$. \\ $\displaystyle{p_{c,D} = \frac{\displaystyle{\sum_{p_{i,t} \in P} p_{i,t}}}{|P|}}$,\\where $P$ is a set of VPMs with $t \in D$ and the longitude/latitude of taxi $i$ belong to $c$ at time point $t$.\end{tabular} \\ \hline
$p_{C,D}$ & \begin{tabular}[c]{@{}c@{}} The average volume of per-minute floating population in a cluster $C$ at time points in $D$ \\ $\displaystyle{p_{C,D} = \frac{\displaystyle\sum_{c \in C} p_{c,D}}{|C|}}$\end{tabular}\\ \hline
$p_{h,c,D}$ & Similar to $p_{c,D}$ but we use only VPMs in $D$ whose hour band is $h$.\\ \hline
$p_{h,C,D}$ & Similar to $p_{C,D}$ but we use only VPMs in $D$ whose hour band is $h$.\\ \hline
ORDINARY & The set of time points before the first case, i.e., the time interval $[t_0, t_f)$\\ \hline
ALERT & The set of time points between the first case and the massive infection case, i.e., the time interval $[t_f, t_m)$\\ \hline
PANIC & The set of time points between the massive infection case and the golden-cross, i.e., the time interval $[t_m, t_g)$\\ \hline
RELIEVED & The set of time points after the golden-cross, i.e., the time interval $[t_g, t_l]$\\ \hline
WORKING & \begin{tabular}[c]{@{}c@{}}Working days in $\mathcal{T}$. We use ORINARY\_WORKING, PANIC\_WORKING, ALERT\_WORKING,\\and PANIC\_WORKING to denote working days in each interval.\end{tabular}\\ \hline
HOLIDAY & \begin{tabular}[c]{@{}c@{}}Holidays in $\mathcal{T}$. We use ORINARY\_HOLIDAY, PANIC\_HOLIDAY, ALERT\_HOLIDAY,\\and PANIC\_HOLIDAY to denote holidays in each interval.\end{tabular}\\ \hline
\end{tabular}}
\end{table}

\subsection{Geographical Unit of Analysis}\label{sec:voronoi}
\begin{figure}[t]
    \centering
    \subfigure[]{\includegraphics[width=0.49\textwidth]{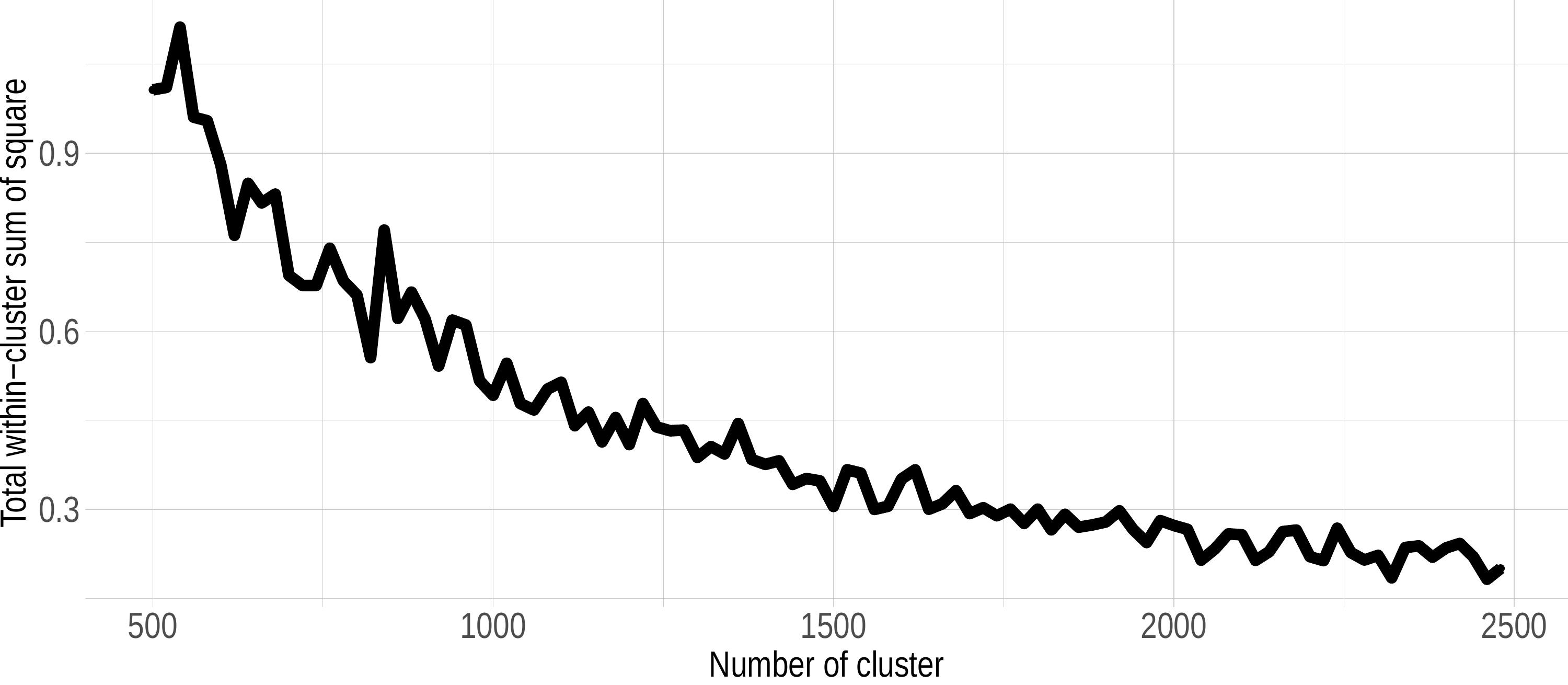}}
    \subfigure[]{\includegraphics[width=0.49\textwidth]{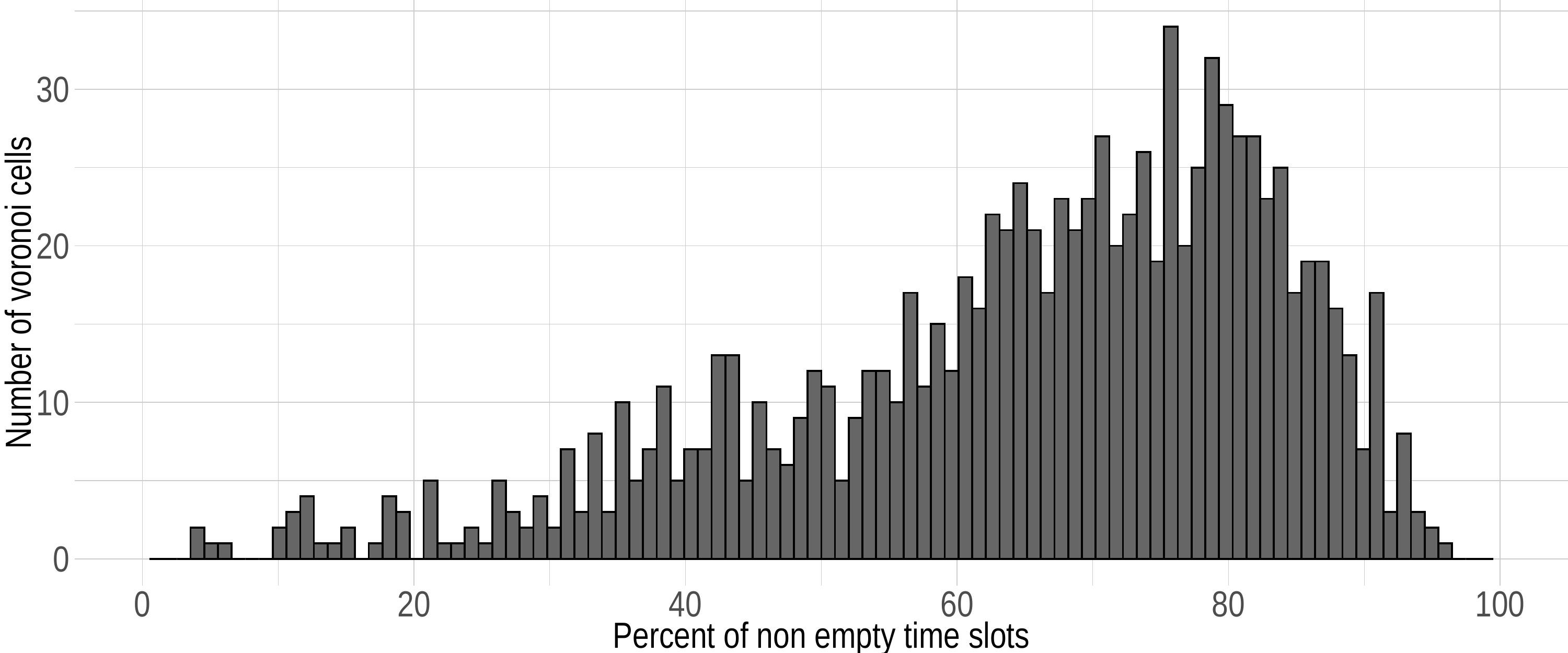}}
    \subfigure[]{\includegraphics[width=0.49\textwidth]{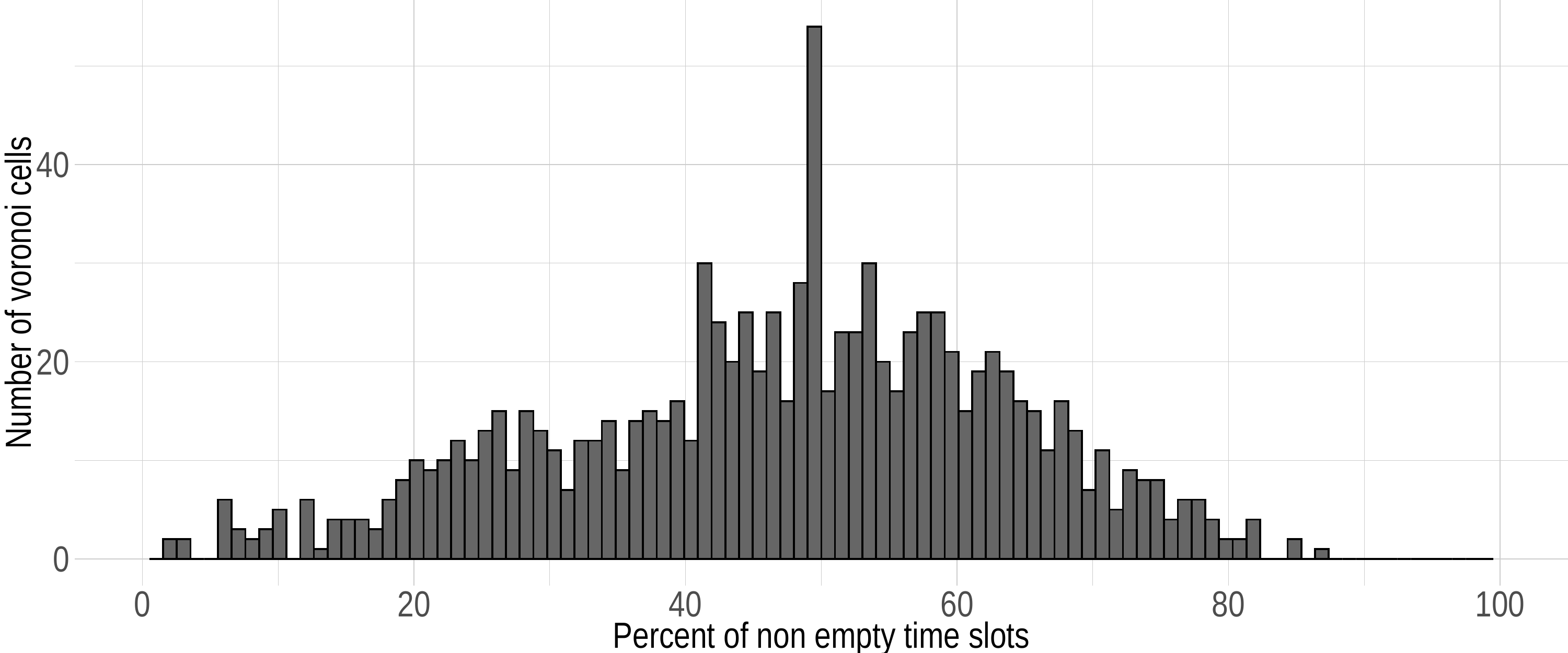}}
    \subfigure[]{\includegraphics[width=0.49\textwidth]{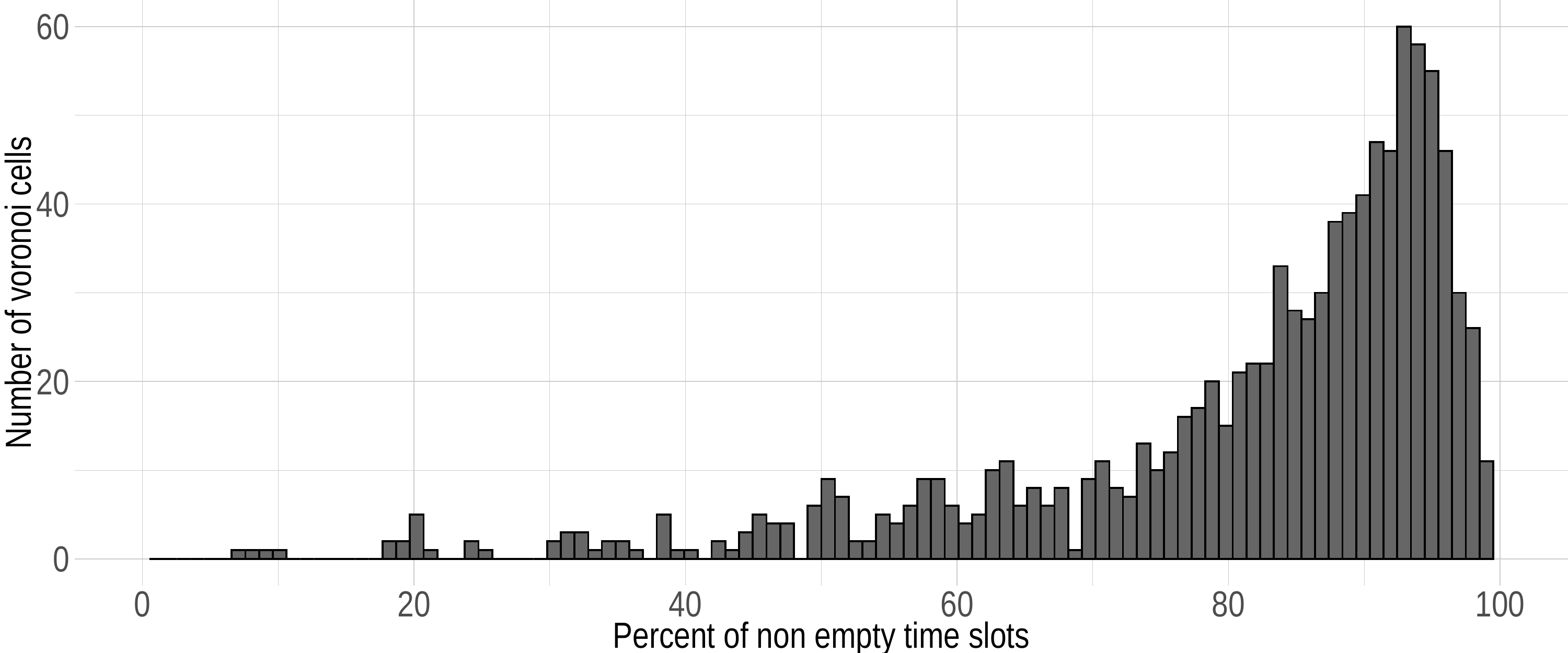}}
    \caption{(a) The $k$-Means clustering of $s_{i,t}$. The X-axis has various $k$ configurations, and Y-axis is the sum of the squared distances to cluster centroids. We choose $k=1,000$ using the elbow method. Note that $k=1,000$ is the starting point of the saturation of the unsupervised clustering evaluation metric. (b) The histogram of the percentage of non-empty time slots when $u$ is one hour. For example, there are about 30 cells whose non-empty percentage is 80\%. (c) The histogram when the unit $u$ is 30 minutes (d) The histogram when the unit $u$ is 2 hours}
    \label{fig:kmeans}
\end{figure}

Although there are municipal boundaries available within the city of Incheon, these administrative boundaries tend to be inconsistent with those of meaningful mobility or people's perception of the neighborhoods, as previous work demonstrates \citep{mckenzie2018identifying,cranshaw2012livehoods}.
In addition, the uneven distribution of taxi locations in our data may cause varying uncertainty levels of mobility across regions if municipal boundaries are used. 
These potential issues lead us to generate geographical boundaries based on taxi locations.
This taxi-report-driven approach to determine the geographical unit of analysis ensures the regularity of taxi density across the city's regions.
One of the best ways to generate geographical boundaries based on location data is Voronoi diagram~\citep{a71ff2df9f3246469efe8638d750498d}.

To generate Voronoi cells, we first run the $k$-Means clustering~\citep{bishop:2006:PRML} for the geo-coordinates of taxis to find $k$ centroids. 
Fig.~\ref{fig:kmeans} (a) shows the sum of the squared distances to cluster centroids, a popular unsupervised metric to evaluate the quality of clustering~\citep{10.1007/s10115-016-1004-2}, as $k$ increases. We use the elbow method~\citep{Thorndike53whobelongs} to choose the optimal $k$ configuration and $k=1,000$ is selected. 
To make use of the elbow method, we first draw a line segment connecting the first and last points (i.e., $k=500$ and $k=2,500$) in Fig.~\ref{fig:kmeans} (a) and calculate the perpendicular distance to the line segment for each point. 
As a result, $k=1,000$ yields the longest perpendicular distance. Using the 1,000 centroids identified by the $k$-Means clustering, we draw a Voronoi diagram and the entire city can be segmented into 1,000 Voronoi cells, as depicted in Fig~\ref{fig:volume} (a).

\subsection{Temporal Unit of Analysis}
As mentioned, $p_{i,t}$ is reported every minute from the partner telecom company, which poses a similar challenge to that of the geospatial unit of analysis but in a temporal manner. 
If the VPM data is used directly, it may not cover all the temporal range for a given Voronoi cell $c$, due to the uneven distribution of taxis across the time slots. 
This means we also need to determine the temporal unit of analysis for covering an enough number of time slots for each Voronoi cell. 
We take the following approach to decide the temporal unit.

First, let $u$, e.g., one hour, be a new temporal unit after aggregating the original time unit $t$. We divide the entire time period $\mathcal{T}$ into many time slots of size $u$, $\{[0,u), [u,2u), \dots\}$. 
We count the number of unique MAC addresses for each time slot $[xu,(x+1)u)$ and then calculate the percentage of the time slots where the count is non-zero in a cell $c$. We choose such a configuration of $u$ where the number of cells with non-empty periods (i.e., the count is non-zero) is large enough. 
Fig.~\ref{fig:kmeans} (b) shows the histogram for the non-empty time slot percentages when $u$ is one hour. In that case, a majority of cells have more than 50\% of non-empty time slots. We think that the histogram of $u = $ 30 minutes has many cells with empty time slots, and $u = $ two hours is too coarse to observe meaningful changes in our analysis. 
Therefore, we use one hour as the temporal unit of analysis.
% because the raw data at the 1-second resolution is biased due to uneven distribution of taxis at a given time. 
% So we aggregated to 1 hour based on sensitivity graph of how temporal resolution is related to the geographical coverage of taxis. 

% (the number of empty cells vs. time-aggregation resolution.

\subsection{Land Use of Voronoi Cells} \label{sec:tag}

Because human mobility patterns depend on the characteristics of the Voronoi cells, mobility needs to be understood by land-use. 
Government-provided land-use exists, but based on our qualitative examination, it does not reflect the land-use of each polygon well, maybe because people's use of the land is often mixed and varies significantly.
For example, the Incheon National Airport area is designated as a green area according to the government's classification, but people's main use is international travel.
This inconsistency between the official land-use classification and their manifestation is common in many cities that present a high population density in Asia \citep{tong1997advantages}. 
This makes it challenging to use the government's land-use classification and necessitates identifying the ground truth of the land-use in each cell in a different way.

For identifying the land-use of Voronoi cells, four student assistants who are familiar with the city of Incheon manually assessed the land-use of each cell based on the amenities and places in it. 
They were asked to rank four land-use types for each Voronoi cell: \emph{residential, commercial, green, and industrial}, instead of selecting the best use because many cells were a mix of different uses. 
For each cell $c$, we randomly assigned three raters and have them rank the land-use types for each cell without talking to each other to minimize peer-influence. 
We classified each cell with a land-use type that was rated by equal to or more than two raters based on the manual coding results.
To evaluate the quality of the initial tagging results, we use the Fleiss' kappa~\citep{gwet2008computing}, denoted $\kappa$, which is a standard way to measure inter-rater agreement on rankings. 
The Fleiss' $\kappa$ was 0.41, which is a moderate agreement level and suggests a reasonable coding result for the analysis. After that, we encouraged people to discuss and make a consensus for each cell's land-use type.

% Need to provide how to generate temporal features and how to conduct clustering of polygons to detect different landuse (figures + graphs needed). Vanessa Frias's paper cited here. 

% Also, polygons were filtered based on volumes? (if a polygon has too low number of individuals, we omit them?)

\subsection{Socio-Political Events related to COVID-19}\label{sec:events}

The first COVID-19 case in South Korea was reported on January 20th, denoted $t_f = $ 00:00:00, 2020-01-20 and the first massive infections started from a cult ceremony in the City of Daegu on February 18th, denoted $t_m = $ 00:00:00, 2020-02-18~\citep{Senn:2009}. 
Even though the City of Daegu was about $250km$ away from Incheon, this massive outbreak brought a serious concern to South Korean people, reminding them of the criticalness of social distancing practices.  
About a month later, since this event, the ``golden-cross,'' where the number of daily new confirmed cases became smaller than that of daily discharged patients from self-quarantine or hospitalization for the first time, happened on March 13th, denoted $t_g = $ 00:00:00, 2020-03-13 \citep{golden_cross_news}. 
This event was meaningful to people, signaling that the COVID-19 outbreak was going downhill. 

These events overall were symbolically noticeable to people because the Korea Centers for Disease Control and Prevention (KCDC) provided guidelines or announcements on these days, and online/offline news media also highlighted these events through many articles~\citep{kcdc:2009}. 
Using these socio-political events that are symbolically and practically meaningful about COVID-19, we divide the entire period $\mathcal{T}$ into four segments: (1) before the first case in Korea, i.e., $[t_0, t_f)$, (2) between the first case in Korea and a large outbreak in the City of Daegu, i.e., $[t_f, t_m)$, (3) between the outbreak in Daegu and the golden-cross, i.e., $[t_m, t_g)$, and (4) on and after the golden-cross, i.e., $[t_g, t_l]$. 
We use $t_0$ and $t_l$ to denote the first and last time points, respectively, in our dataset, i.e., $t_0 = $ 00:00:00, 2019-12-18 and $t_l = $ 23:59:59, 2020-04-30. 
These four time-segments are characterized as ORDINARY, ALERT, PANIC, and RELIEVED, respectively, based on how South Korean people feel about COVID-19. 
We also note that ALERT includes the Lunar New Year (January 25th), which is one of the biggest holidays in east Asia, and RELIEVED includes the national legislative elections day (April 15th).

% also newspapers highlighted these days significantly (citations needed for the newspaper pages and policy pages for the particular dates). 

%%%%%%%%%%%%%%%% Myeong
\subsection{Socio-Economic Status}\label{sec:imputation}

Because we focus on Voronoi cells within the city of Incheon, high-resolution data that indicates the socio-economic status of residents is needed at the cell level to answer RQ2. 
However, South Korea has not made the socio-economic status data open to the public at the granular geographical level within a city; instead, it only provides income and education status data at the city and province level \citep{kostat}.
Due to a lack of official socio-economic status data at the Voronoi cell level, we use the housing transaction data as a proxy for measuring the socio-economic status. This is feasible because, in South Korea, real estate occupies the most significant portion of wealth, especially since the 2000s \citep{kim2018dissonance}.

Although housing prices weakly predicted the household income in the 1990s in South Korea \citep{kim1998socioeconomic}, the structure of the real estate financing market has radically changed. 
Since the 2000s, real estate owners have heavily relied on the \emph{Chonsei} system, a unique rent system in South Korea that allows the owner to keep a large amount of money from tenants during their rental period, instead of receiving monthly rent \citep{kim2018dissonance}.
This system made it possible for the owners to finance by themselves in acquiring additional real-estates, without relying on banking systems such as a mortgage. 
As a result, real estate ownership underpinned by the Chonsei system has become a significant source of socio-economic inequality in South Korea in the 2010s by hampering the entrance of new owners to the real estate market, because the institutional support for real-estate financing was insufficient for newbies \citep{kim2018dissonance}.
These studies on the Korean real-estate market justify that the housing price data based on real-world transactions can be used as a proxy of socio-economic status for the Voronoi cells.

The housing price data was part of the real-estate transaction dataset for the period between April 1st in 2019 and April 30th in 2020.
This data was released by Ministry of Land, Infrastructure and Transport through their open data portal \citep{mlit}. 
While the transaction data provides precise locations with physical addresses, the real-estates' locations cover only part of the Voronoi cells. 
This led us to use a geospatial imputation method to estimate the housing prices of empty cells. 
When the locations of the real estate transactions were geographically aggregated into Voronoi cells, 573 out of 1000 cells had housing price values.
For each of the empty cells, we selected the real-estates that are located within a $1km$ radius from the centroid of the cell and imputed the housing price by averaging the unit prices of the selected real-estates. 
The length of the radius, $1 km$, was determined based on the size of ``Dong,'' the smallest census unit in South Korea. 
The mean value of all Dong areas in the city of Incheon is $3.4 km^2$, and the median value of the areas is $2.1 km^2$, showing a long-tale distribution.  
This means the average radius of Dong areas is about $1.04 km$, and the median radius of Dong areas is about $0.82 km$, which justifies $1 km$ as the radius for imputing the housing prices of empty cells.

After the imputations with the $1 km$ threshold, 114 Voronoi cells (11.4\% of entire cells) were still empty, meaning that there were no real-estate transactions within a $1 km$ radius from the centroids of these cells. 
Because many of these cells were located in the peripheral areas of the city or green areas such as mountains, we used a well-known statistical imputation method, \emph{mean substitution} for these regions: the 114 cells were imputed with the average value of all the other cells. 
We use the results of this multi-step imputation process as the proxy of socio-economic status for the Voronoi cells.

\section{Results}
In this section, we describe our main analysis results, including their detailed charts. After describing descriptive statistics in our dataset, we answer all the research questions.

%%%%%%%%%%%%%%%@ Seongkyu
\subsection{Descriptive Statistics}

% Need numbers here with tables. The average number of taxis in each polygon per hour, the change of the number of taxis over time, the average number of individuals in each polygon by time segments, etc. 

% A table of variables needed. Some analysis on assumptions like multivariate normality or variable distributions could be mentioned.

\begin{figure}[t]
    \centering
    \subfigure[]{\includegraphics[width=0.52\textwidth]{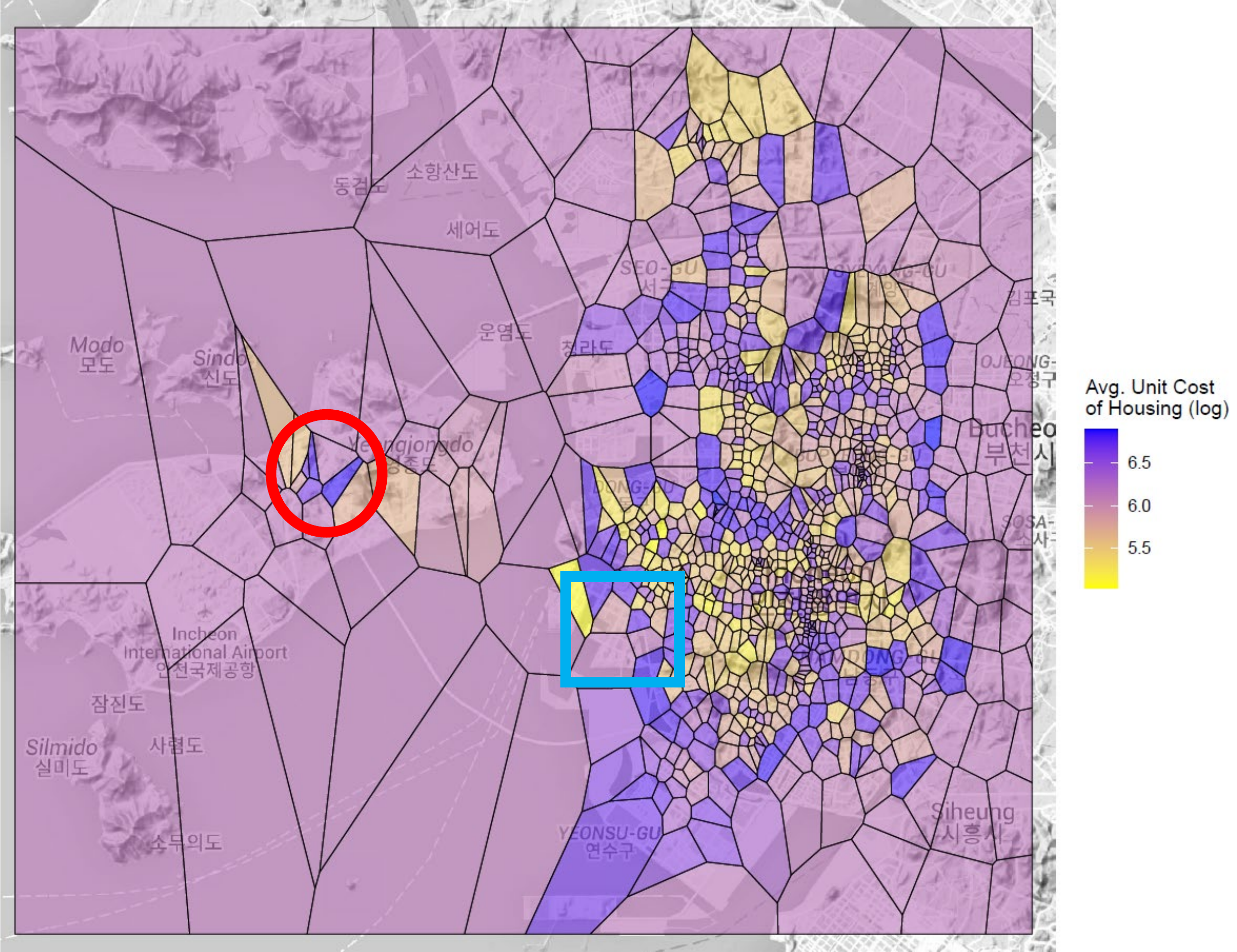}}
    \subfigure[]{\includegraphics[width=0.46\textwidth]{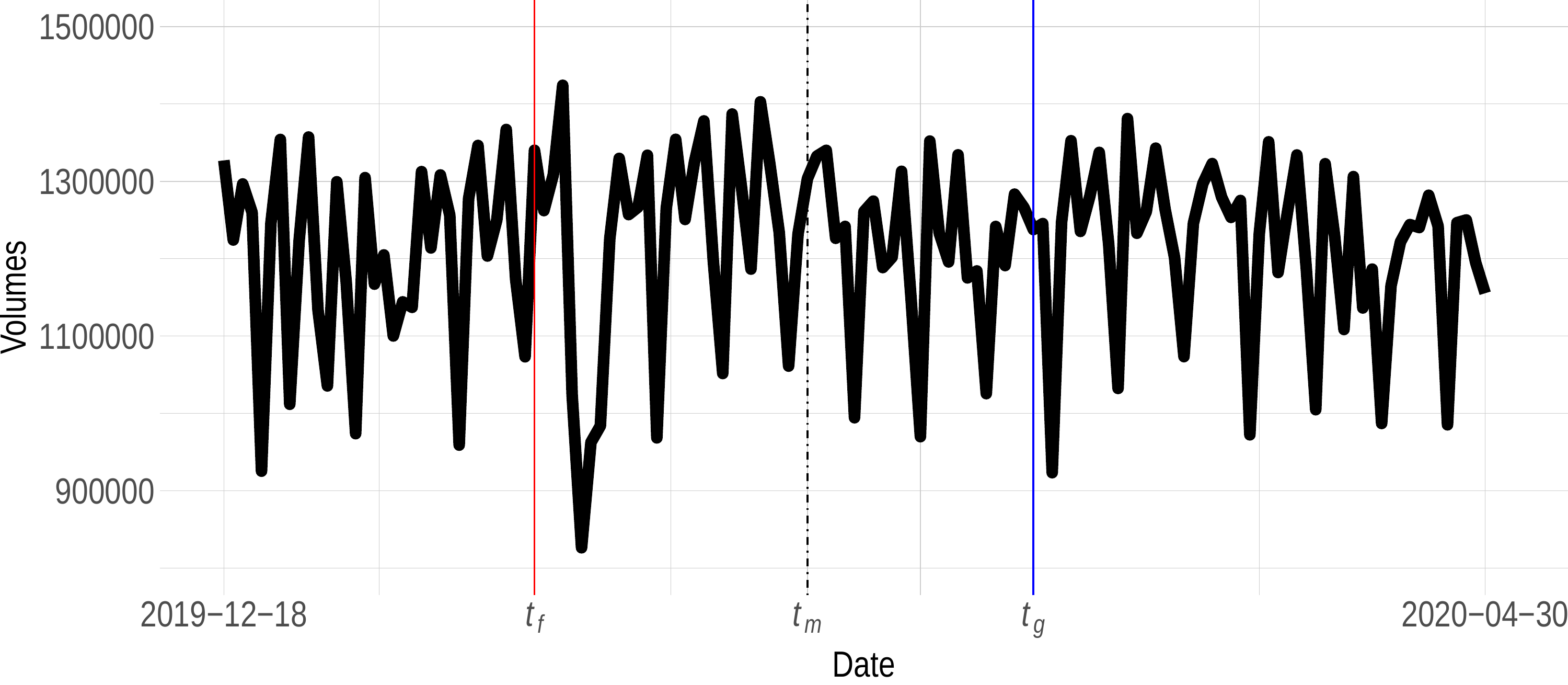}}
    \caption{(a) The average unit prices for housing in the Voronoi cells (after geo-imputation) (b) The total volume of the floating population in Voronoi cells, i.e., the number of unique MAC addresses discovered daily. Note that the volume slightly decreases after the first massive infections at $t_m$. Because the nationwide lock-down had not been imposed in South Korea, the degree of mobility reduction is not as high as that in other countries.}
    \label{fig:volume}
\end{figure}

Fig.~\ref{fig:volume} (a) shows the Voronoi diagram with 1,000 cells generated based on taxi locations in the city of Incheon, following the steps presented in Section~\ref{sec:voronoi}. 
Using the average unit price of the real-estates that were purchased or sold or the value that was imputed based on the method presented in Section~\ref{sec:imputation}, the Voronoi cells are color-coded in Fig.~\ref{fig:volume} (a).
The smallest cell has a size of $0.01 km^2$, the largest has $124 km^2$, and the median size of the cells is $0.2 km^2$, showing a long-tale distribution, just like that of `Dong's.\footnote{The size of the largest cell is $124 km^2$ because it covers the region outside of the city due to the characteristic of the Voronoi diagram. Because the size of the cells is not used in the analysis, this bias that exists on the periphery of the city is negligible.}

The total volume change of the floating population in our dataset is shown in Fig.~\ref{fig:volume} (b). More than 1M unique MAC addresses are collected in a day for the data collection period. 
Compared to that in ALERT, the total volume of working days decreased by 3.3\% ($p$=3.56e-18) during the PANIC period.
We expect that the weak implementation of a social distancing measure in South Korea resulted in a small decrease compared to that in other countries \citep{Beaubien_2020}.

\subsection{RQ1: How did socio-political events about COVID-19 affect human mobility in South Korea?}

To precisely analyze the human mobility influenced by important socio-political events, we first perform a t-test to test whether a Voronoi cell $c$'s mobility is influenced by the occurrence of the first massive infections at $t_m$. 
After dividing each working day and each holiday in ALERT/PANIC into 24 hour bands, respectively, we construct an array of 48 values (24 for working days and 24 for holidays), denoted as follows:
\begin{align}\begin{split}
    S_{c,ALERT\_WORKING}=&\{p_{[0,1),c,ALERT\_WORKING}, p_{[1,2),c,ALERT\_WORKING}, \dots\},\\
    S_{c,PANIC\_WORKING}=&\{p_{[0,1),c,PANIC\_WORKING}, p_{[1,2),c,PANIC\_WORKING}, \dots\},\\
    S_{c,ALERT\_HOLIDAY}=&\{p_{[0,1),c,ALERT\_HOLIDAY}, p_{[1,2),c,ALERT\_HOLIDAY}, \dots\},\\
    S_{c,PANIC\_HOLIDAY}=&\{p_{[0,1),c,PANIC\_HOLIDAY}, p_{[1,2),c,PANIC\_HOLIDAY}, \dots\},
\end{split}\end{align} where $p_{-,c,-}$ means the average of all the VPMs in each hour band in cell $c$ during all working days/holidays of ALERT/PANIC.

We use $S_{c,ALERT\_WORKING}$ and $S_{c,ALERT\_HOLIDAY}$ (resp. $S_{c,PANIC\_WORKING}$ and $S_{c,PANIC\_HOLIDAY}$) to construct an array with 48 values representing the mobility signature of $c$ during ALERT (resp. PANIC). We compare these two arrays, one for ALERT and the other for PANIC, using a paired t-test in each cell $c$. 
The massive infections at $t_m$ put people into a panic so comparing ALERT and PANIC is an effective method to understand the mobility change during COVID-19. 
Using the p-value threshold of 0.05, we found that 146 cells presented decreases in mobility, 47 cells showed increases, and the other remaining cells had no significant changes in mobility before and after $t_m$. 
We analyze the decreasing, increasing, and non-changing cases separately to highlight each case's unique patterns as well as to identify diverse populations. 

\begin{figure}[t]
    \centering
    \subfigure[Residential cluster (82 Voronoi cells)]{\includegraphics[width=0.49\textwidth]{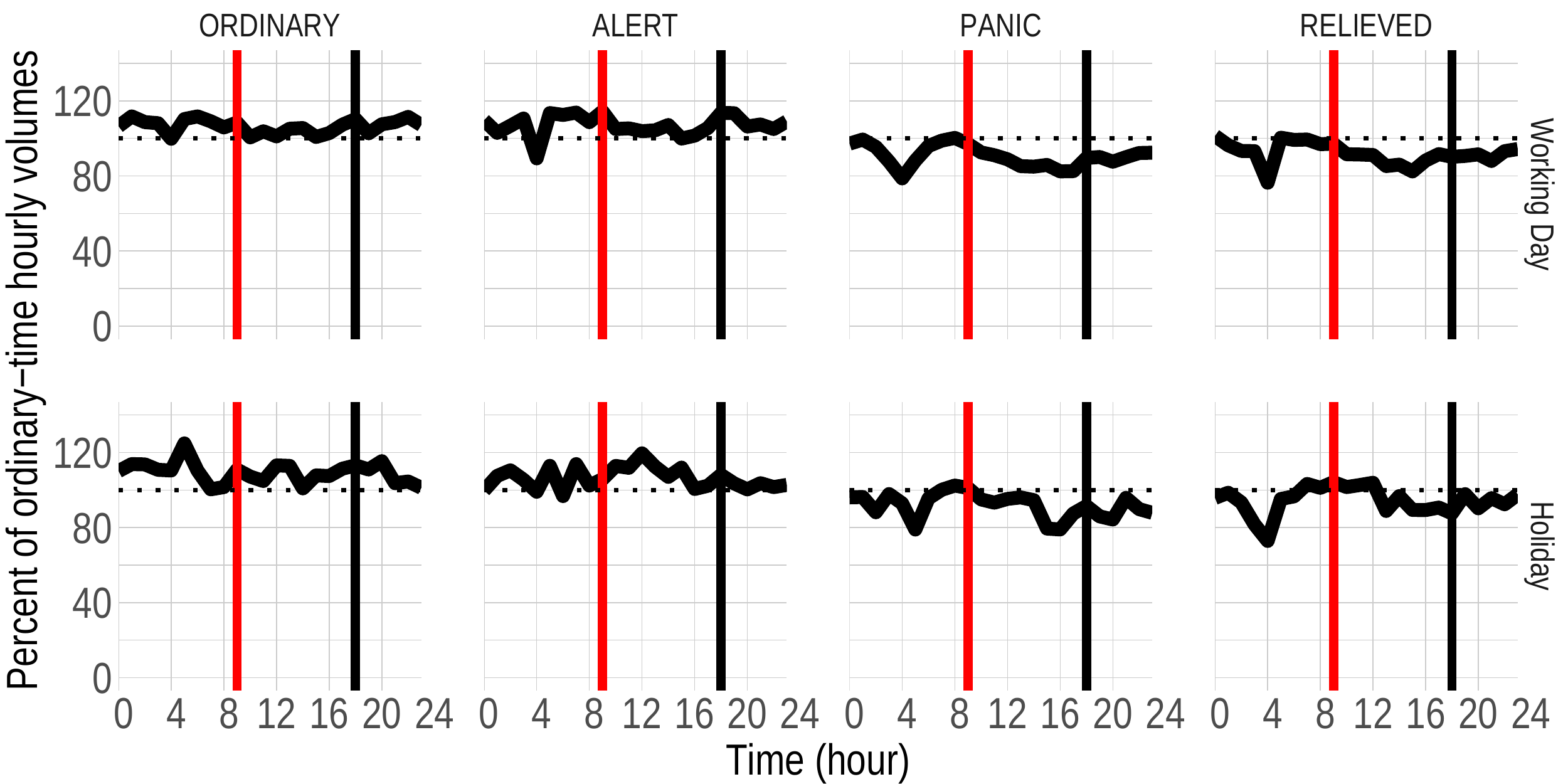}}
    \subfigure[Commercial cluster (37 Voronoi cells)]{\includegraphics[width=0.49\textwidth]{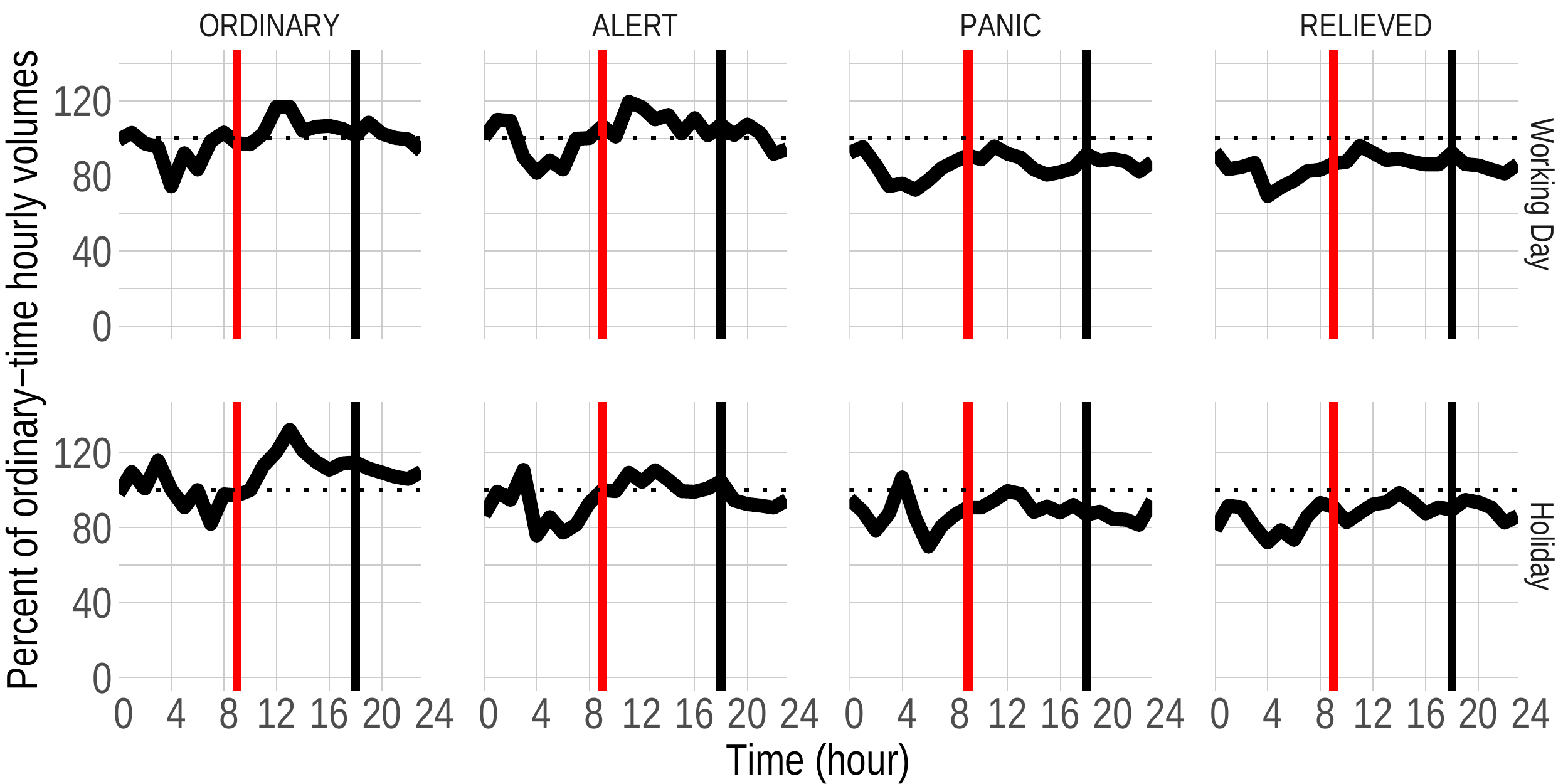}}
    \subfigure[Industrial cluster (9 Voronoi cells)]{\includegraphics[width=0.49\textwidth]{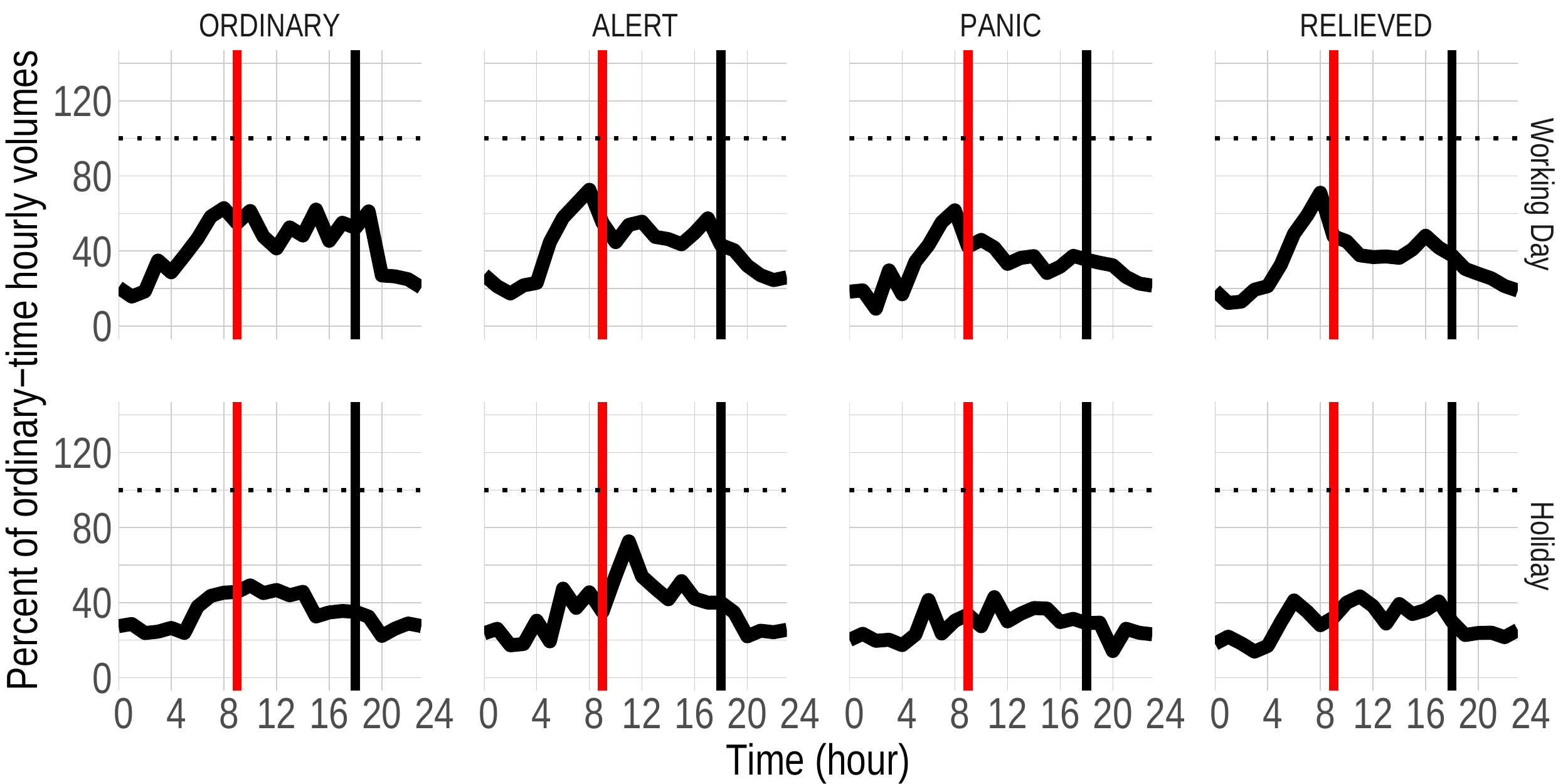}}
    \subfigure[Green cluster (13 Voronoi cells)]{\includegraphics[width=0.49\textwidth]{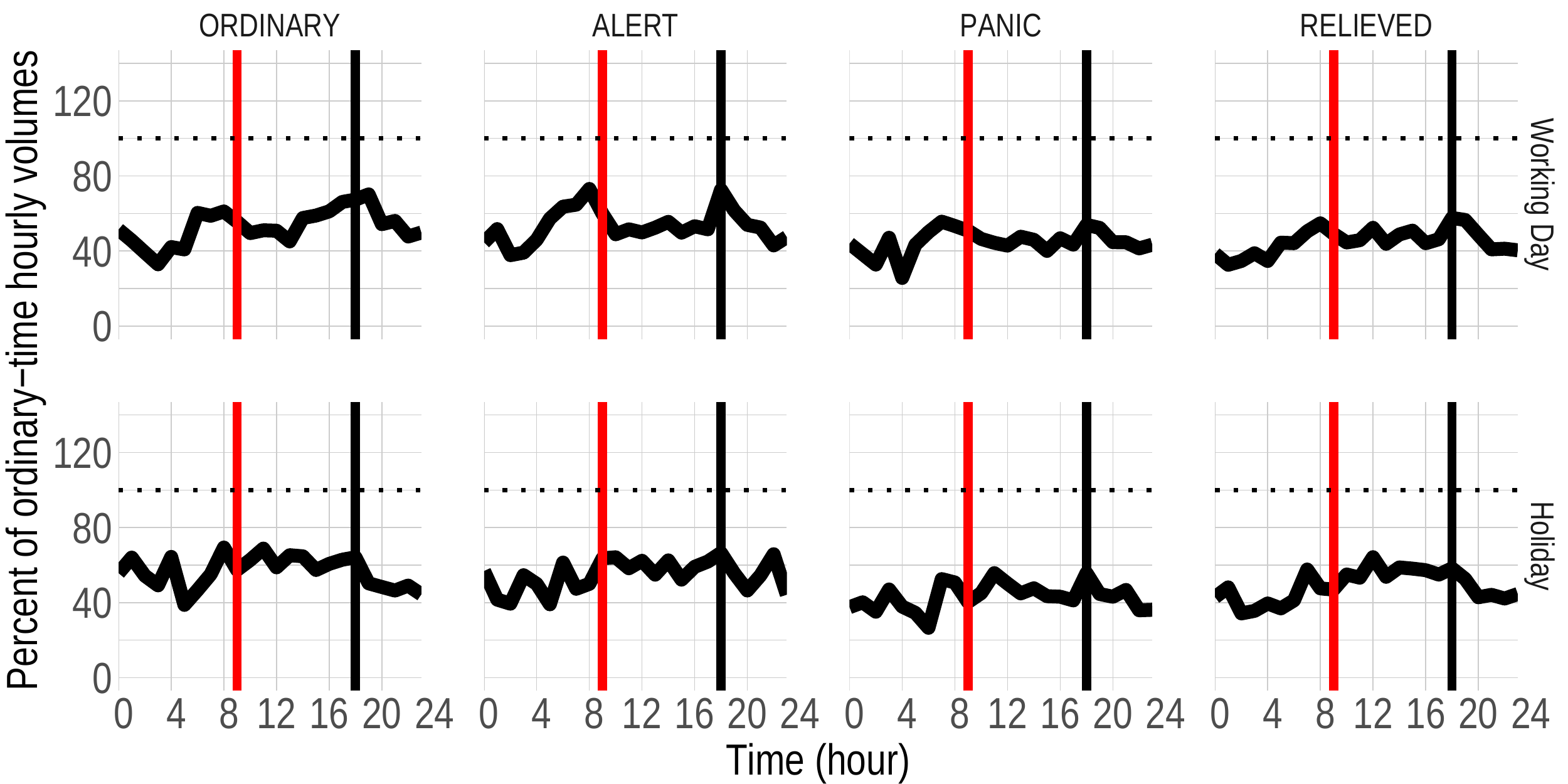}}
    \caption{The hourly mobility patterns for various land-use types of the Voronoi cells where human mobility decreased after $t_m$. The vertical red and black lines mean 9am and 6pm, respectively.}
    \label{fig:rq1}
\end{figure}

We classify Voronoi cells with decreased mobility into four types of land-uses, namely, \emph{residential}, \emph{commercial}, \emph{industrial}, and \emph{green}, using the manual tagging, as described in Section~\ref{sec:tag}. 
Fig.~\ref{fig:rq1} shows their mobility patterns. 
$Y$-axis is the volume relative to the hourly baseline volume of ORDINARY.\footnote{100\% means volume at an hour band is same as the mean volume in the same hour band of ORDINARY.} 
Both the residential and commercial regions show significant drops in mobility in PANIC. 
The cells for residential use present a mobility ratio of -14.50\% ($p$=3.81e-39) between ALERT and PANIC --- these residential cells include many standalone high-rise apartments with commercial places on their lower floors, which are usually more expensive than other residential types. The commercial cluster that consists of commercial places around bus stations, stadiums, administrative buildings presented -13.41\% ($p$=2.39e-38) between ALERT and PANIC.
In particular, the working day mobility of the residential regions decreases by -16.00\% ($p$=8.68e-78). % of 8.68748987e-78}
Their mobility is not recovered even in RELIEVED.
For the industrial cluster around the port of Incheon, the mobility does not decrease during working days.
However, weekend mobility shows a meaningful difference between ALERT and PANIC by -25.71\% ($p$=1.32e-6). 
In the 12 Voronoi cells of the green cluster, their mobility decreases during PANIC by -16.13\% ($p$= 1.88e-15). 
These cells include not only park areas but also hills and small mountains, so their absolute volume of mobility is much smaller than the baseline.

One interesting pattern across residential, commercial, and industrial clusters is that while morning time mobility decreases in general, mobility at the peak rush hour time around 8 AM does not decrease that much. 
Also, slight peaks around the 6 PM (rush hours) during ORDINARY working days disappear in PANIC. 
This suggests that many people who can work remotely started working from home, while people who need to go to work still commute to workplaces but may have changed their commuting patterns to avoid rush hours. 
Similarly, mobility peaks during ORDINARY disappear in both holidays and working days for the green cluster. 
This means the massive outbreak at $t_m$ affected people's commuting behavior as well as their leisure activities during weekends. 

% The mobility change from PANIC to RELIEVED is slightly smaller than that of ORDINARY 15.89\% ($p$=2.72e-5). % calculated in total, need to change if only for weekend

\begin{figure}[t]
    \centering
    \subfigure[Residential cluster (29 Voronoi cells)]{\includegraphics[width=0.49\textwidth]{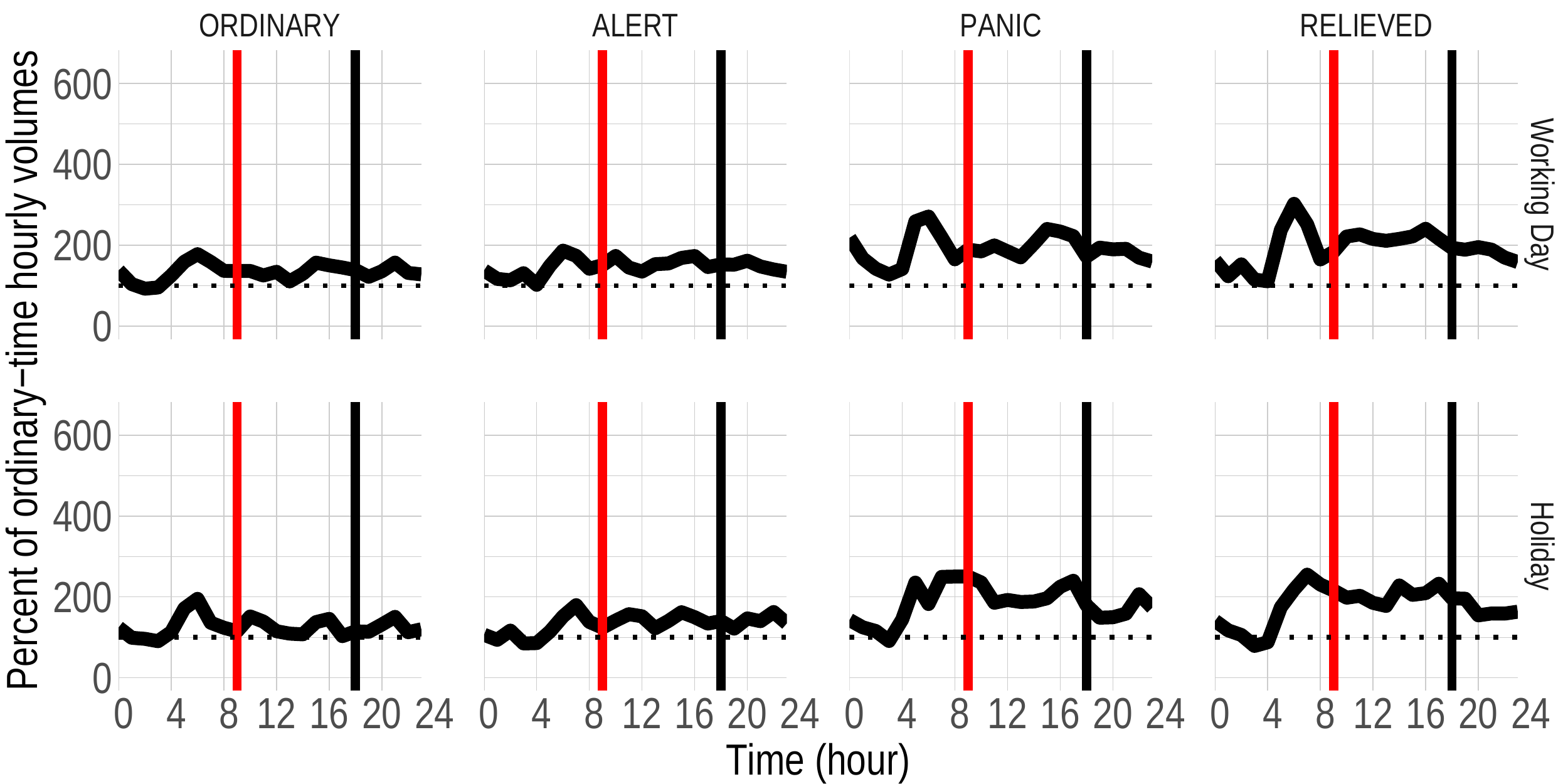}}
    \subfigure[Commercial cluster (13 Voronoi cells)]{\includegraphics[width=0.49\textwidth]{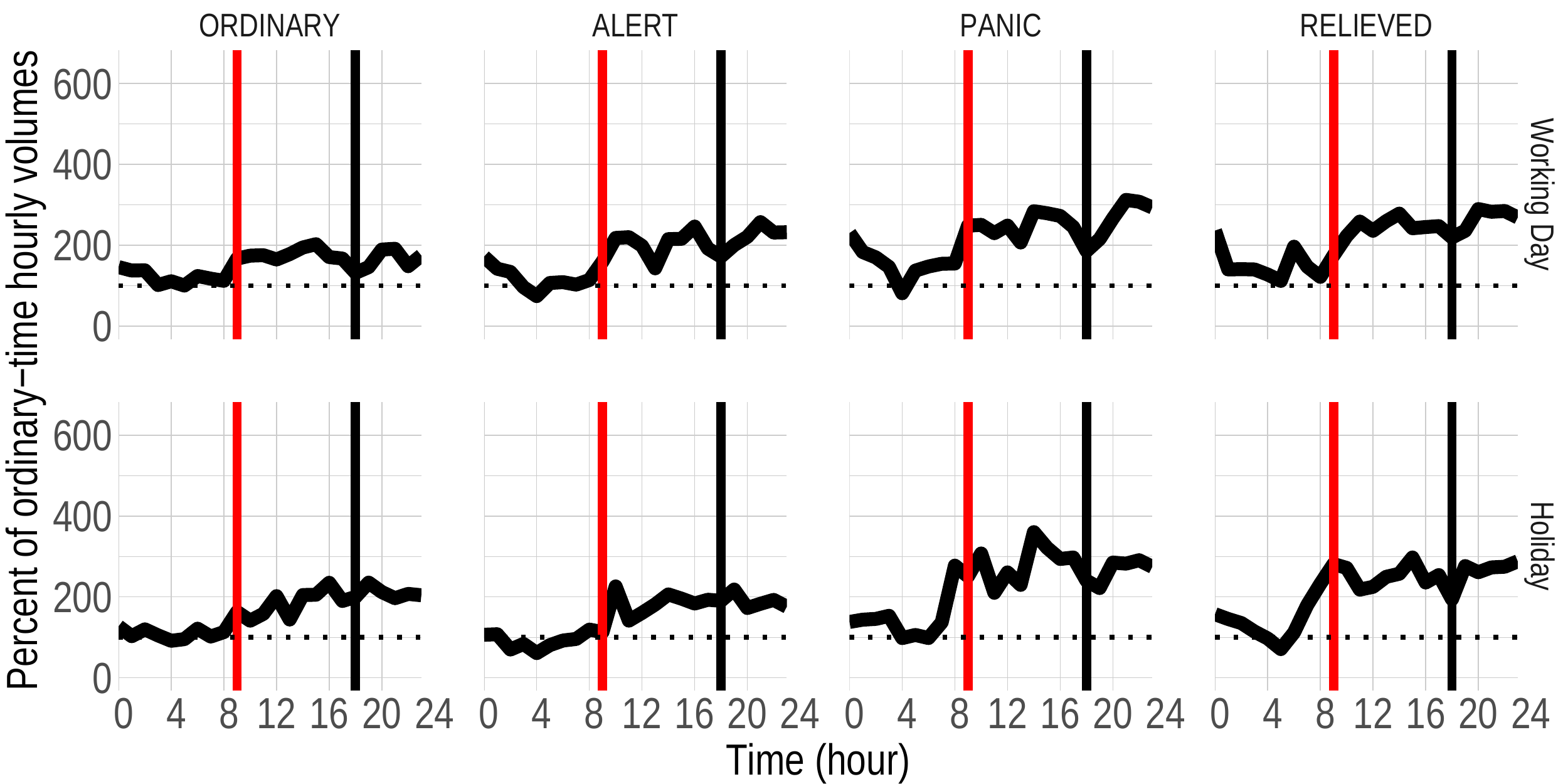}}
    \subfigure[Industrial cluster (3 Voronoi cells)]{\includegraphics[width=0.49\textwidth]{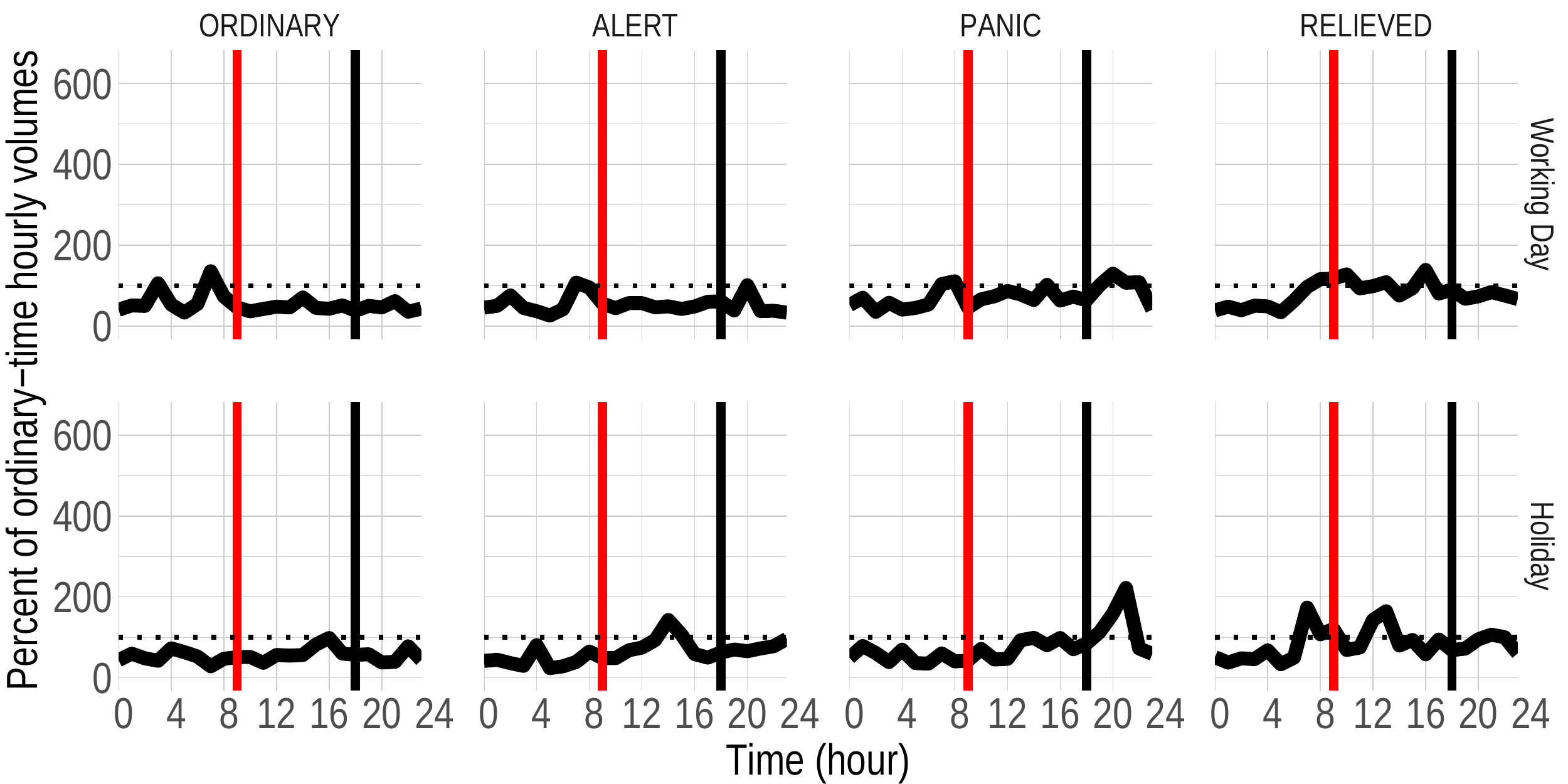}}
    \subfigure[Green cluster (1 Voronoi cell)]{\includegraphics[width=0.49\textwidth]{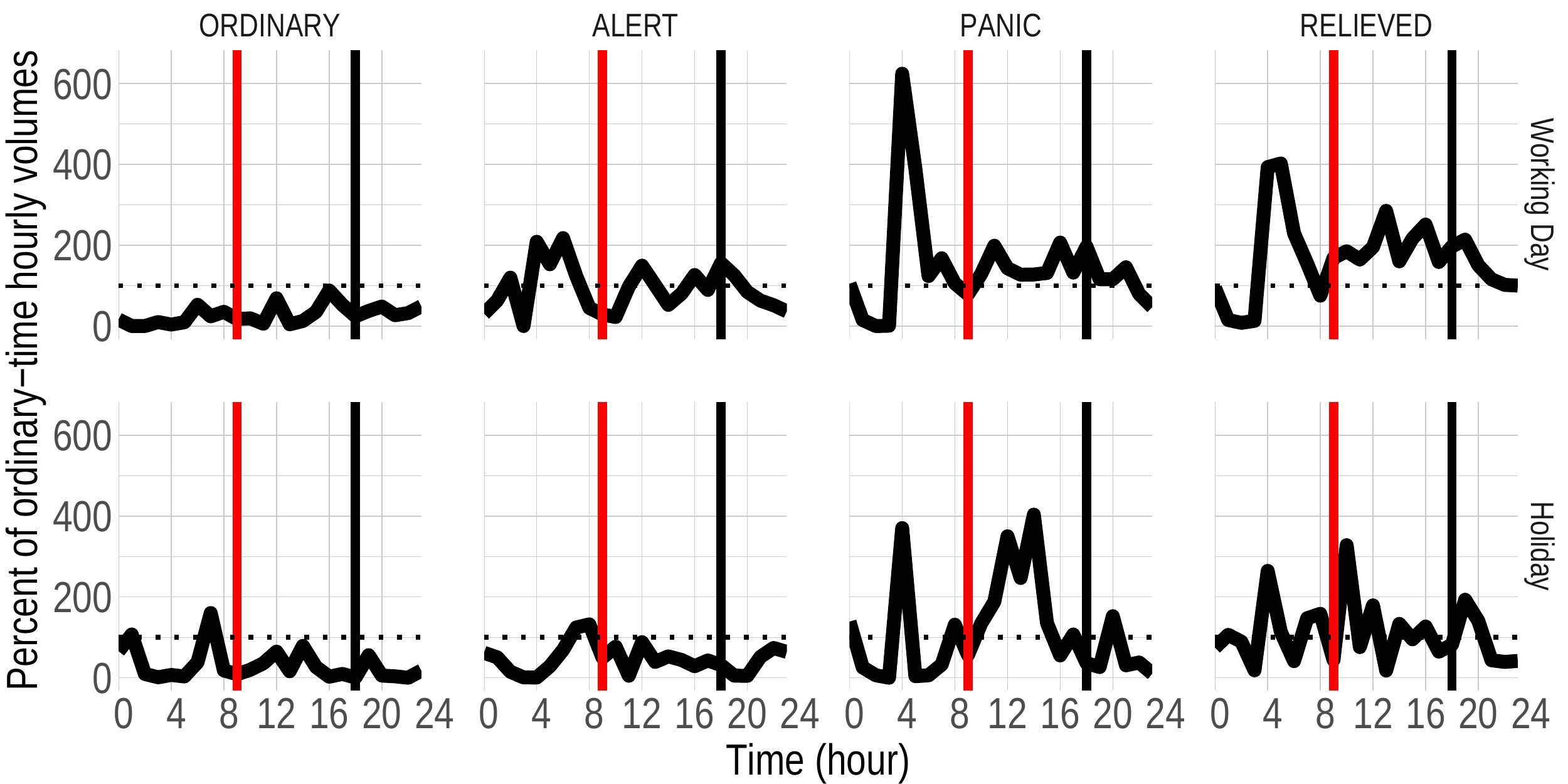}}
    \caption{The hourly mobility patterns for various land-use types of Voronoi cells where human mobility increased after $t_m$.}
    \label{fig:rq1-2}
\end{figure}

\begin{figure}[t]
    \centering
    \subfigure[Residential cluster (146 Voronoi cells)]{\includegraphics[width=0.49\textwidth]{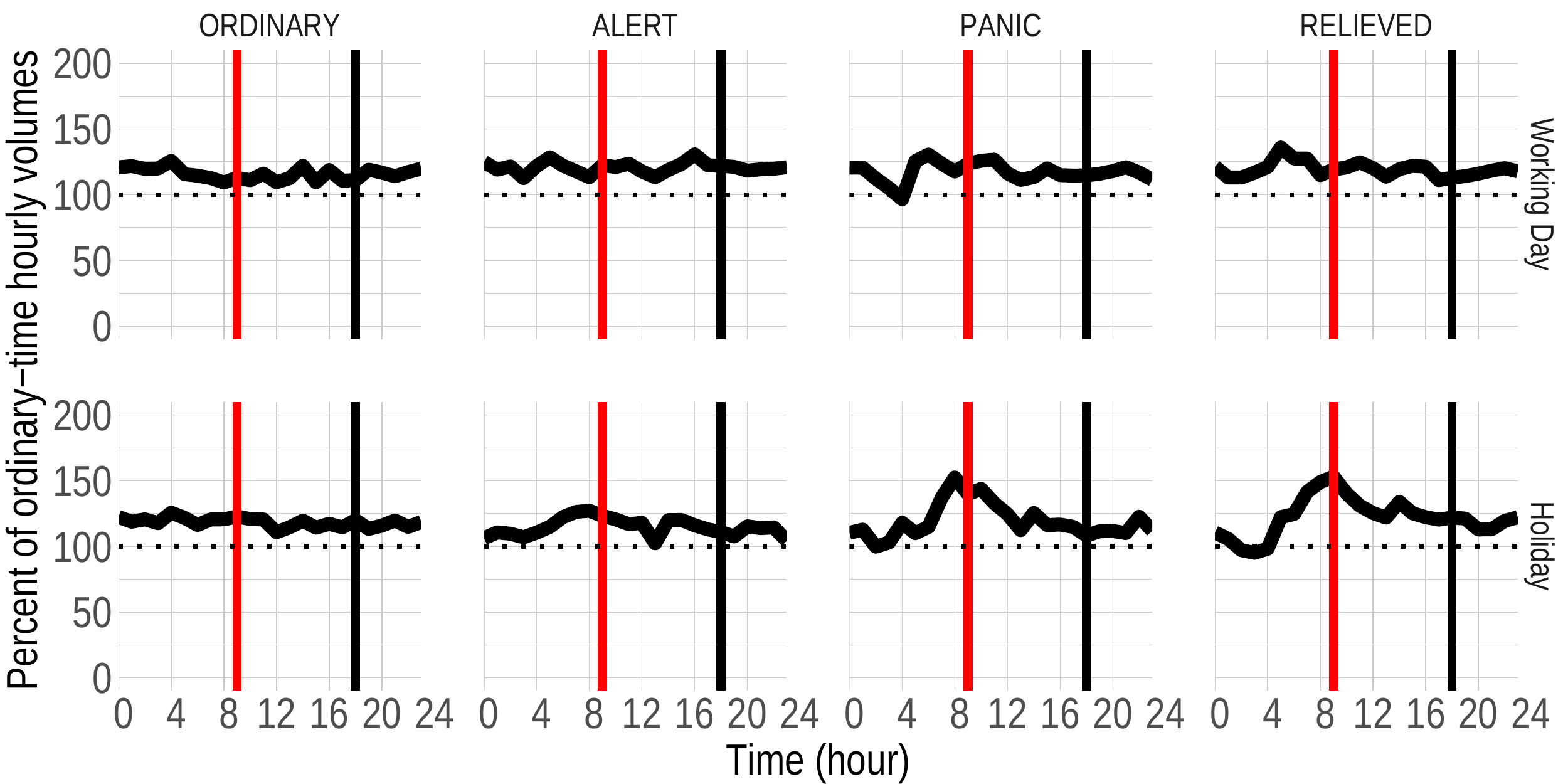}}
    \subfigure[Commercial cluster (36 Voronoi cells)]{\includegraphics[width=0.49\textwidth]{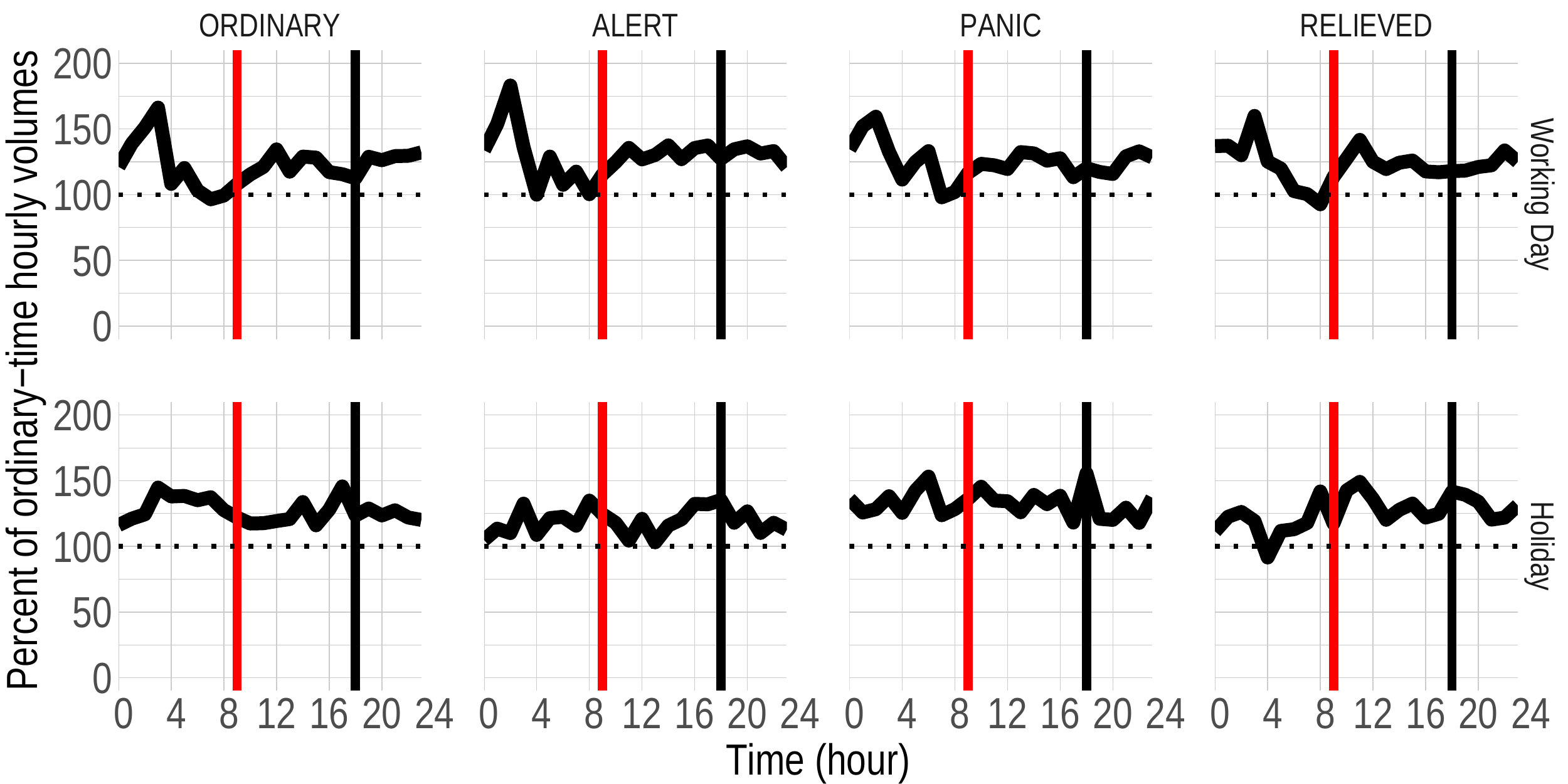}}
    \subfigure[Industrial cluster (18 Voronoi cells)]{\includegraphics[width=0.49\textwidth]{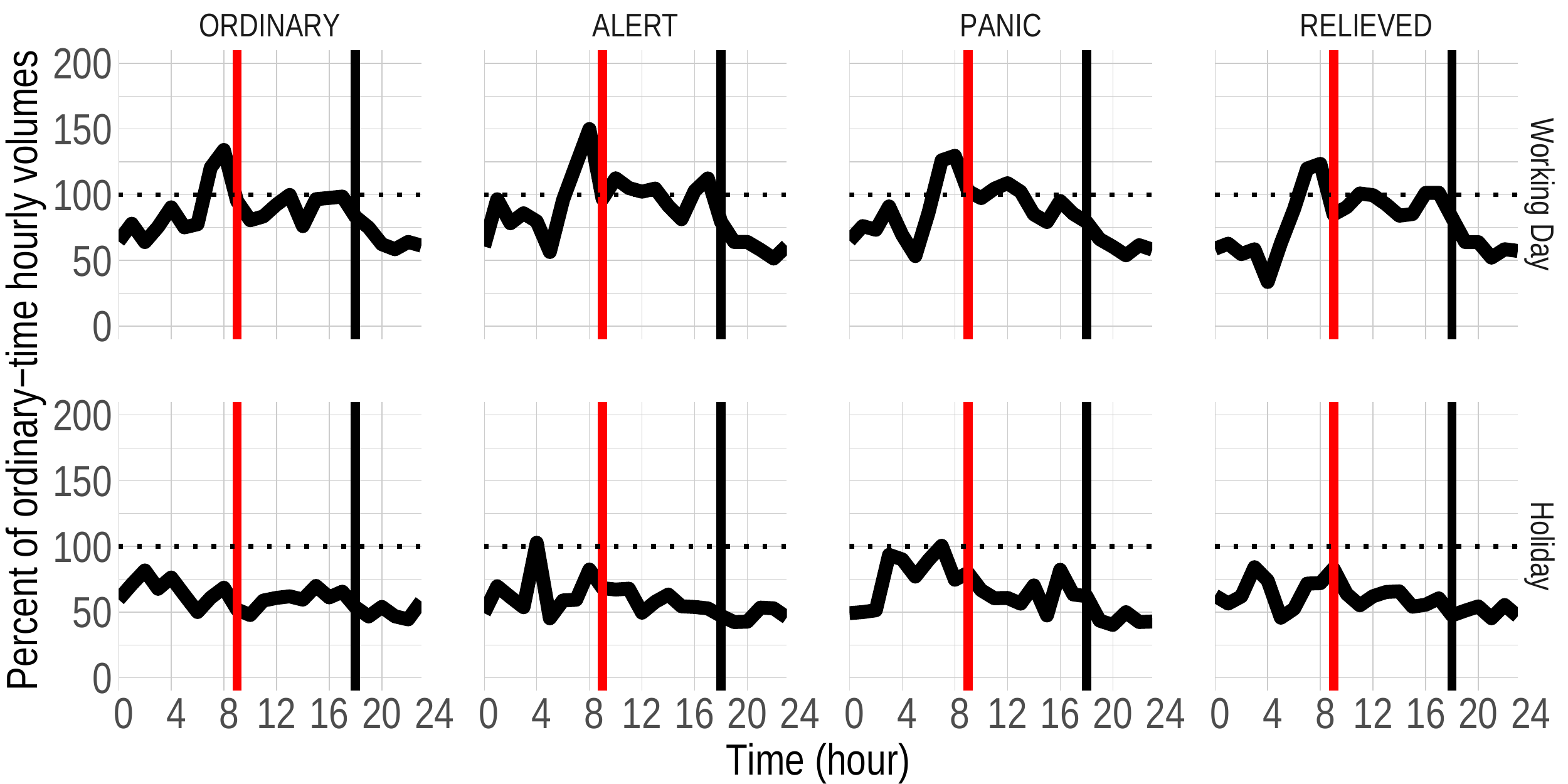}}
    \subfigure[Green cluster (10 Voronoi cell)]{\includegraphics[width=0.49\textwidth]{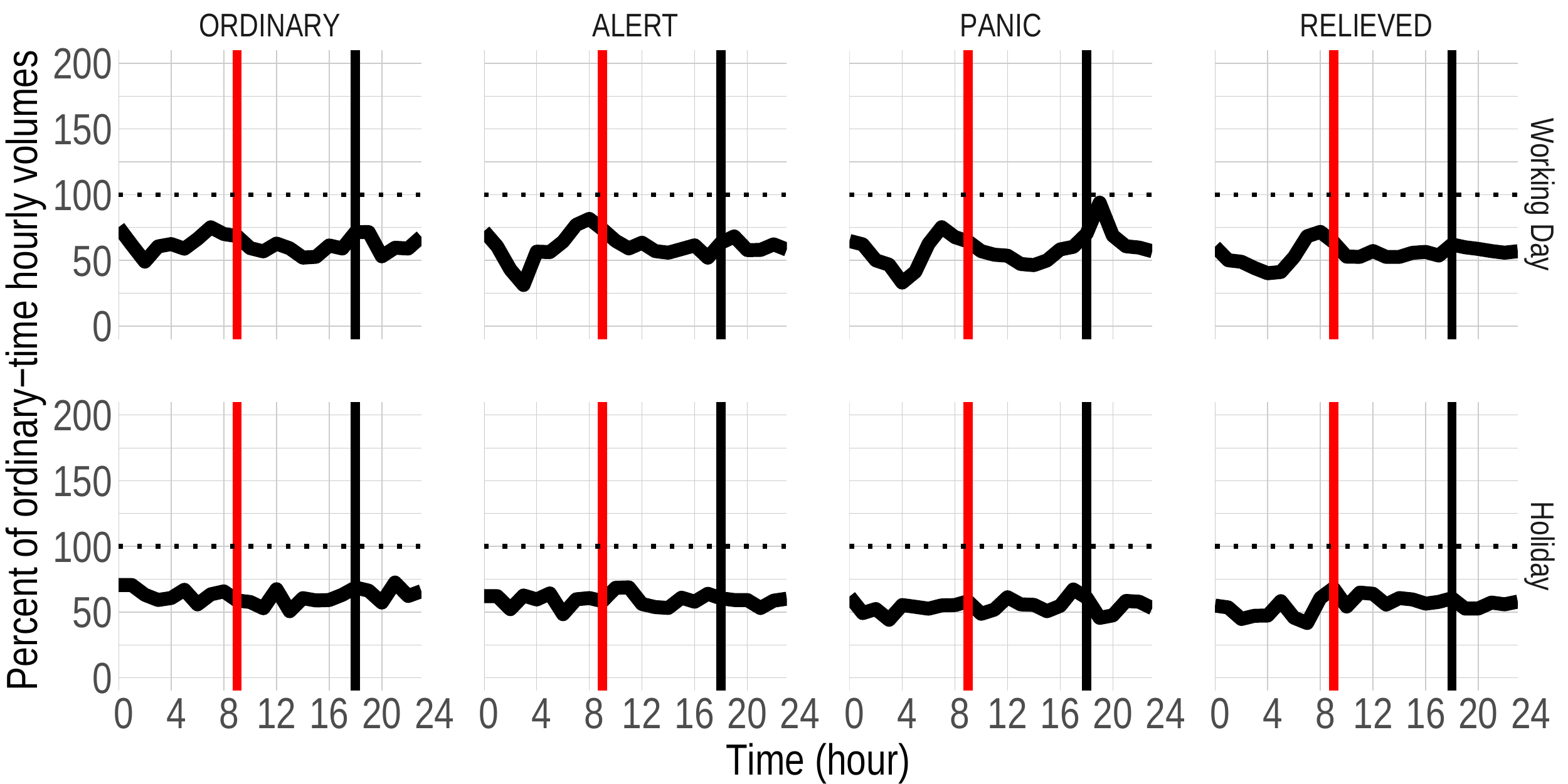}}
    \caption{The hourly mobility patterns for various land-use types of Voronoi cells where human mobility does not change after $t_m$.}
    \label{fig:rq1-3}
\end{figure}

Fig.~\ref{fig:rq1-2} shows the mobility patterns of the increasing cases. In all but the green cluster, their increasing rates are 80.14\% ($p$=1.76e-5) --- this residential cluster has many apartment complexes, a typical living type for many people in South Korea, for the middle and working classes. In particular, the mobility of the green cluster in PANIC increases the most in our dataset compared to that in ALERT for early morning on working days (245.99\%, $p$ = 0.029). We speculate that this radical increase is due to people's increased exercise in early morning times under the COVID-19 situation. For holidays, both early-morning and afternoon mobility increases significantly, maybe, also due to outdoor exercise (166.30\%, $p$=0.015).
This cell that shows such an abnormal pattern is a park and the only cell in the green cluster. 
This region is highlighted with a red circle in Fig.~\ref{fig:volume} (a), and its nearby cells are one of the most expensive areas and the best school district in the city of Incheon.\footnote{The reputation of the school district is a key determinant that shapes housing prices in South Korea. In this particular cell, there are two high-profile high schools special-purpose for science/international subjects and highly competitive.} 

Another interesting pattern is that there are new peaks observed in early morning times and late afternoons on working days during PANIC in the residential cluster. 
In the commercial cluster, the pattern is similar to the residential cluster where the morning and afternoon peaks are slightly shifted to earlier times, respectively, during working days. 
These changes might be due to commuting citizens who had to shift their commuting times to avoid the crowds.
Also, the increased volumes on both working days and holidays in these clusters indicate that some people had work more during and after the pandemic, which might be caused by decreased income.

Finally, Fig.~\ref{fig:rq1-3} presents Voronoi cells with non-significant mobility changes (residential: -0.62\%, $p=$0.198, commercial: -0.34\%, $p=$0.713, industrial: -0.84\%, $p=$0.418, green: 0.68\%, $p=$0.438). 
As noted, $p$-values are not significant in these cells between ALERT and PANIC, when the paired t-test is used. 
However, we can still observe some pattern changes for particular time frames.
For example, on working days in the green cluster during PANIC, a peak in mobility is observed at around 6 PM, maybe due to people's outdoor exercises. 
During the morning times on holidays in the residential cluster, a peak is observed as well, maybe due to extra work or outdoor activities. 
These cells are a random sample of 210 cells out of all the 777 cells that do not present changes in mobility because many cells are in peripheral areas and we assumed that a random sampling of these cells would show a representative mobility pattern of the population (i.e., the 777 cells).

% \paragraph{Comparison between the decreasing and increasing cases: } 
% Therefore, we think the people living near the park are rich and have more chances to take care of themselves than others. \NP{It is the case for the residential and commercial clusters as well. The residential and commercial clusters in Fig.~\ref{fig:rq1-2} consist of more expensive cells than those in Fig.~\ref{fig:rq1}. See the next subsection for more detailed analyses about socio-economic factors.}

% need graphs and tables of entire patterns here. 
% (how the entire volume changed over time, particularly before and after important dates)
% (need ANOVA tables here -- if there's a systematic difference between the periods. How much $R^2$ and F?)

% How about after golden-cross?

% The patterns need to presented (1) as a whole, (2) by landuse, and (3) showcasing particular polygons such as a big station.

%%%%%%%%%%%%%%%@ Myeong
\subsection{RQ2: How did human mobility in South Korea vary depending on socio-economic status?}

\begin{figure}[t]
    \centering
    \subfigure[The top quartile of housing price]{\includegraphics[width=0.49\textwidth]{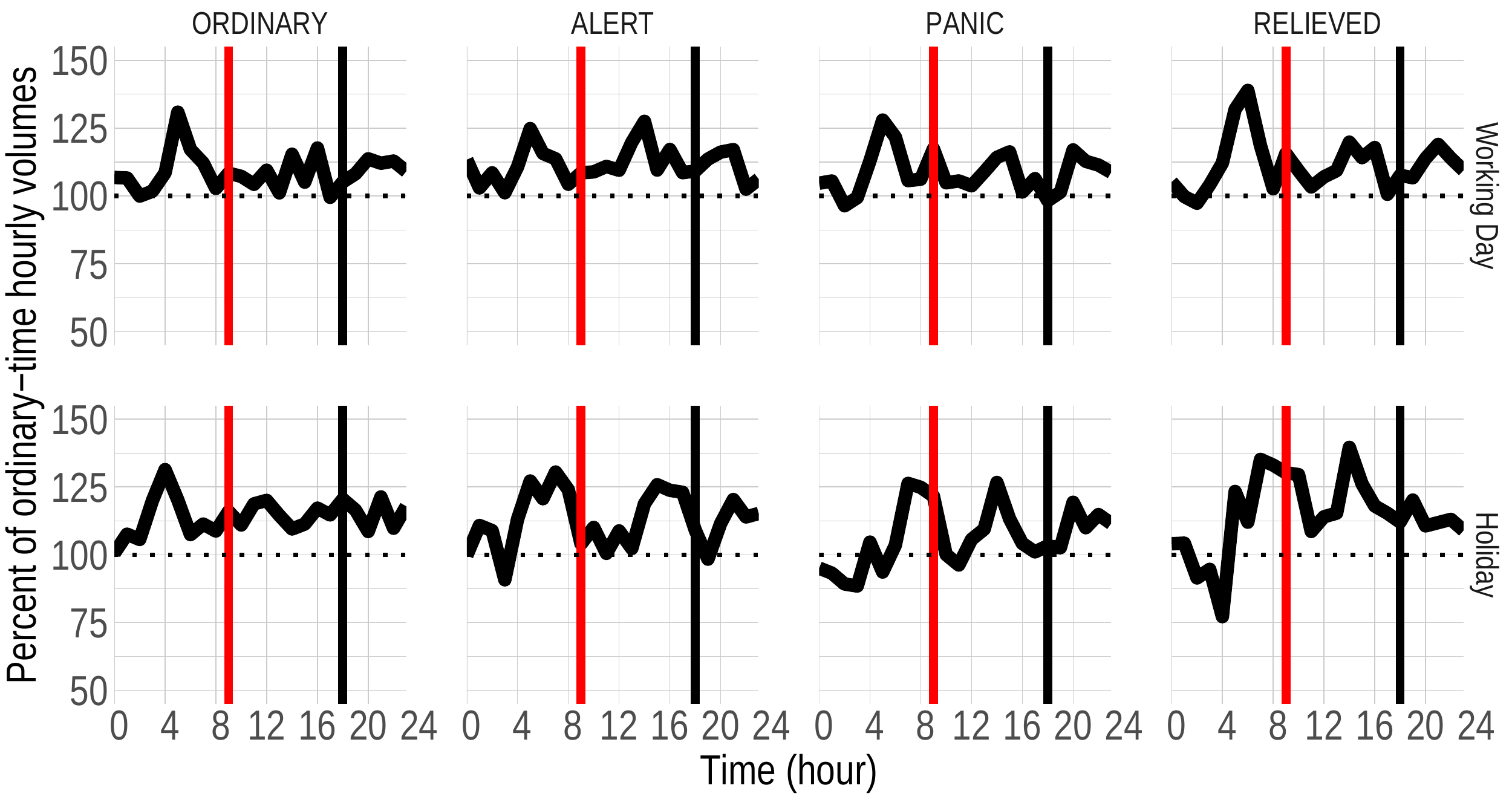}}
    % \subfigure[Commercial cluster (40 Voronoi cells)]{\includegraphics[width=0.49\textwidth]{images/top_25_to_50_percent.pdf}}
    % \subfigure[Industrial cluster (9 Voronoi cells)]{\includegraphics[width=0.49\textwidth]{images/top_50_to_75_percent.pdf}}
    \subfigure[The bottom quartile of housing price]{\includegraphics[width=0.49\textwidth]{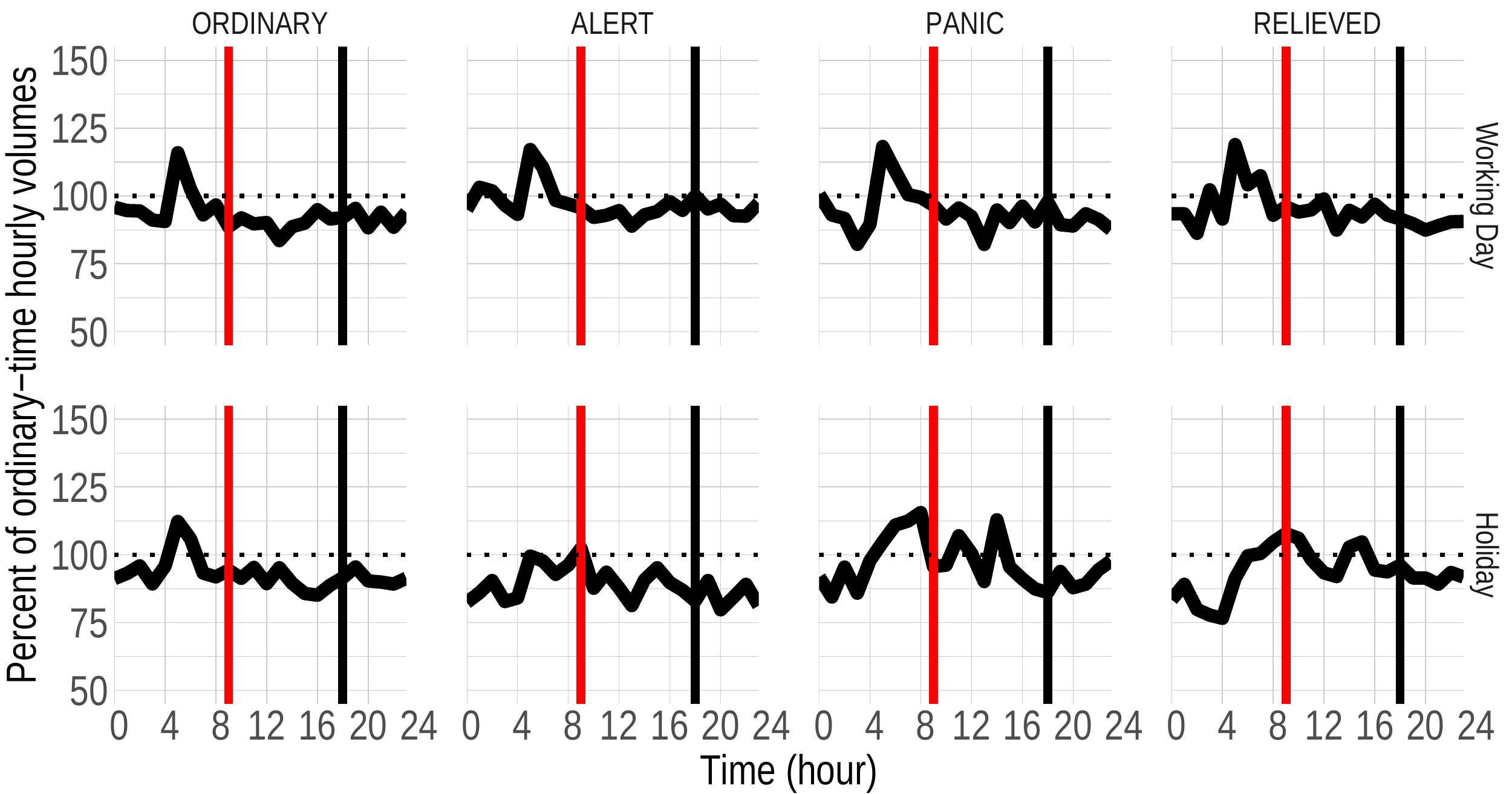}}
    \caption{The hourly mobility patterns for the top and bottom quantiles of the average unit prices for housing.}
    \label{fig:rq2}
\end{figure}

In the aggregate level, there was no significant change in mobility that is shaped by socio-economic differences, which is consistent with the reporting from the Sweden case \citep{dahlberg2020effects}. 
We further analyze the impact of socio-economic status to see how they vary in more contextual conditions.
Using the housing price information in Section~\ref{sec:imputation}, we choose residential cells and group the cells into quantiles. The mobility patterns for the top and bottom quantiles of housing prices are shown in Fig.~\ref{fig:rq2}.
In the top quantile cells in Fig.~\ref{fig:rq2} (a), the average mobility on working days and holidays slightly decreases during the daytime of PANIC compared to that of ALERT (-2.69\%, $p$=0.050). RELIEVED, however, shows an increased mobility pattern by 3.44\% ($p$=0.007) in comparison with PANIC. 
One more interesting point is that the top quantile's mobility is mostly larger than the baseline marked with the dotted horizontal line (i.e., y=100). 
This might be because wealthy areas in South Korea tend to have well-managed parks, and people would have better access to these parks.

In the bottom quantile cells in Fig.~\ref{fig:rq2} (b), there are no significant changes observed in working-day mobility during PANIC, and its mobility pattern is mostly lower than the baseline. 
One possible interpretation is that people still need to work even in the COVID-19 situation, so there are no significant working-day mobility changes. 
Maybe for a similar reason, the golden-cross does not cause any mobility change for the bottom quantile.
Also, increased mobility in the morning and afternoon times on holidays during PANIC compared to ALERT confirms such inferences. 
Meanwhile, its holiday mobility shows meaningful increases across time frames (9.10\%, $p$=1.24e-4). 
However, the absolute volume is not as large as that of the top quantile, maybe due to the limited access to parks and recreational zones.

\begin{figure}[t]
    \centering
    % \subfigure[The hourly mobility of female in 20s]{\includegraphics[width=0.49\textwidth]{images/age_20_gender_F.pdf}}
    % \subfigure[The hourly mobility of male in 20s]{\includegraphics[width=0.49\textwidth]{images/age_20_gender_M.pdf}}
    \subfigure[The hourly mobility of female over 60s]{\includegraphics[width=0.49\textwidth]{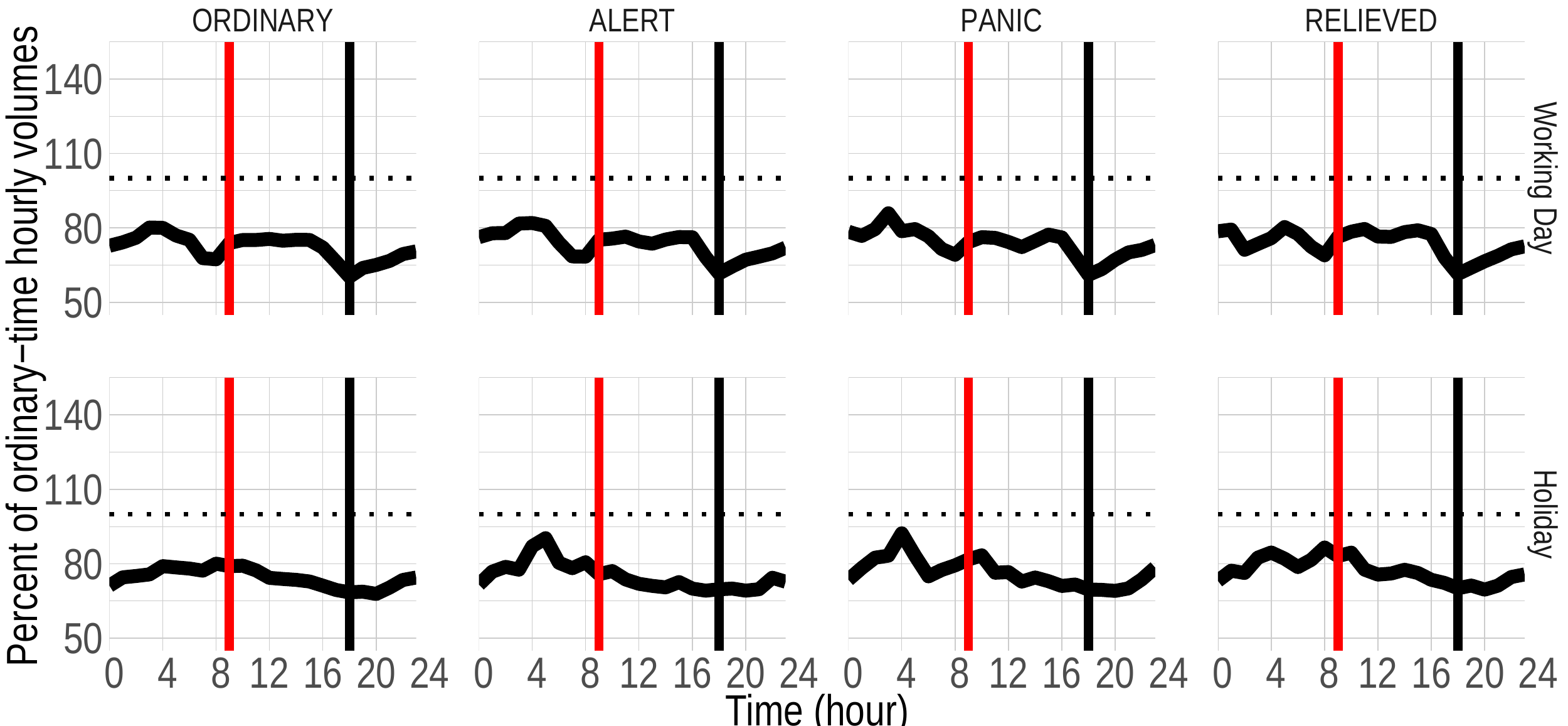}}
    \subfigure[The hourly mobility of male over 60s]{\includegraphics[width=0.49\textwidth]{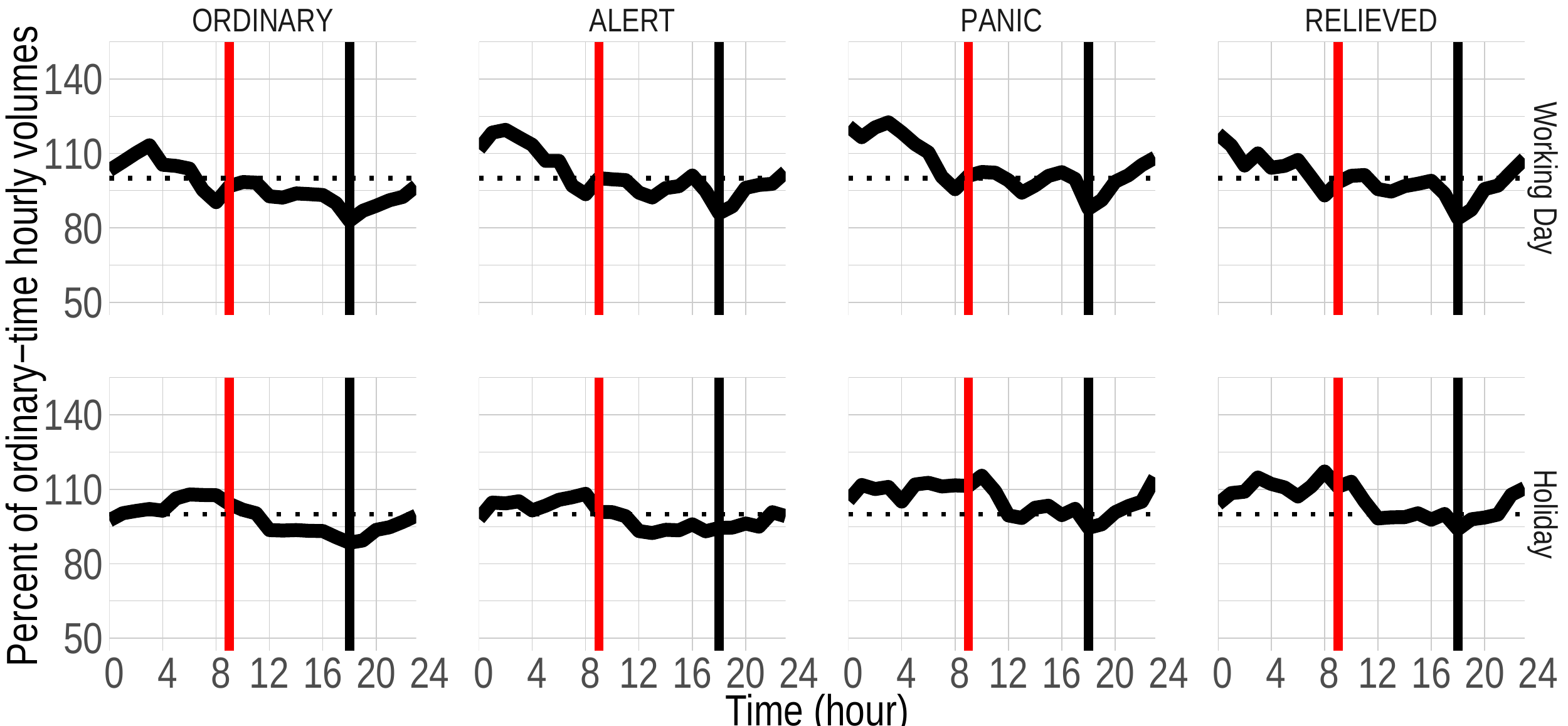}}
    \subfigure[The hourly mobility of female in their 20s]{\includegraphics[width=0.49\textwidth]{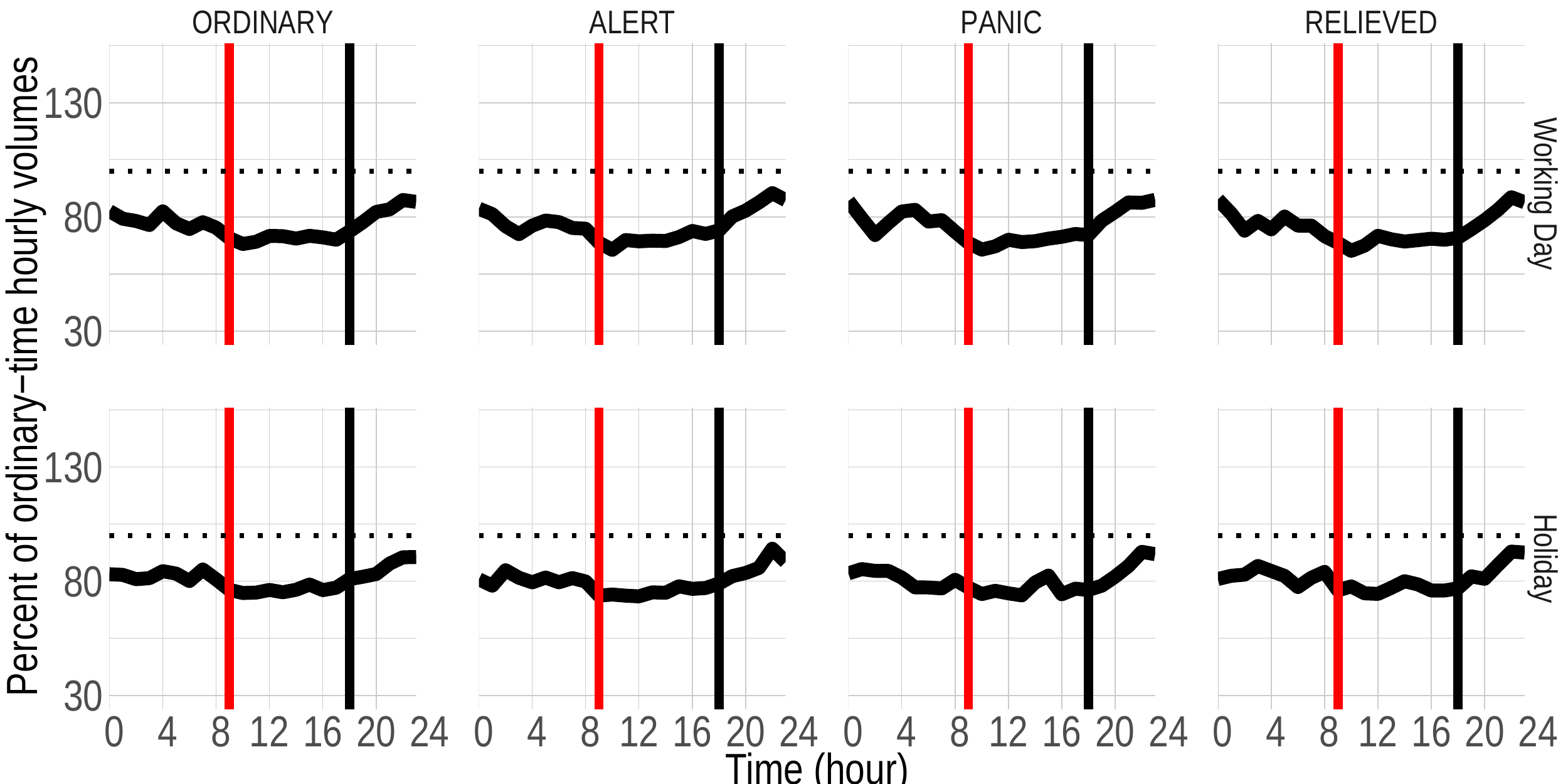}}
    \subfigure[The hourly mobility of male in their 20s]{\includegraphics[width=0.49\textwidth]{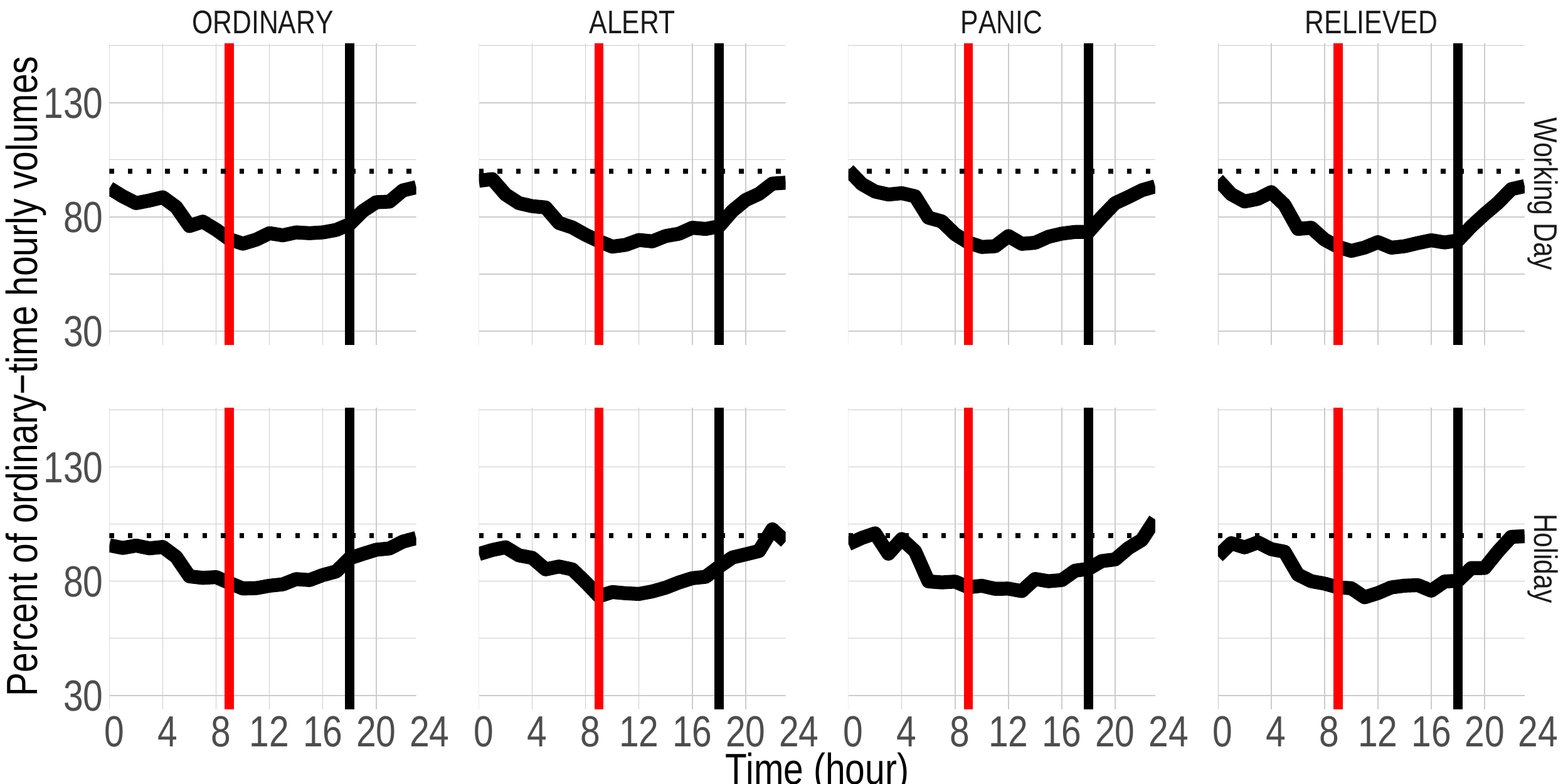}}
    \subfigure[The mobility change by age in the residential cluster with decreased mobility from Fig.~\ref{fig:rq1} (a)]{\includegraphics[width=0.49\textwidth]{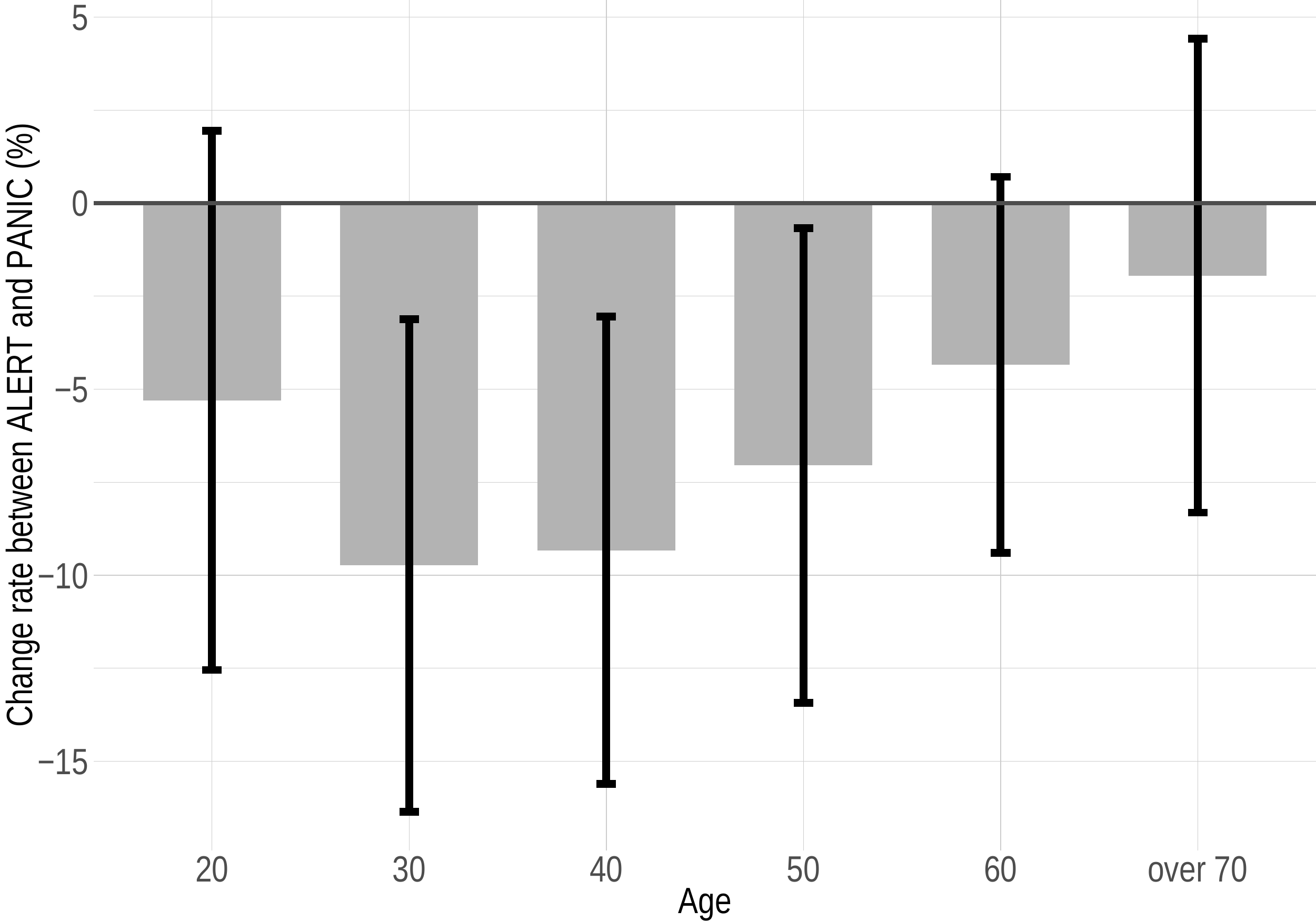}}
    \subfigure[The mobility change by age in the residential cluster with increased mobility from Fig.~\ref{fig:rq1-2} (a)]{\includegraphics[width=0.49\textwidth]{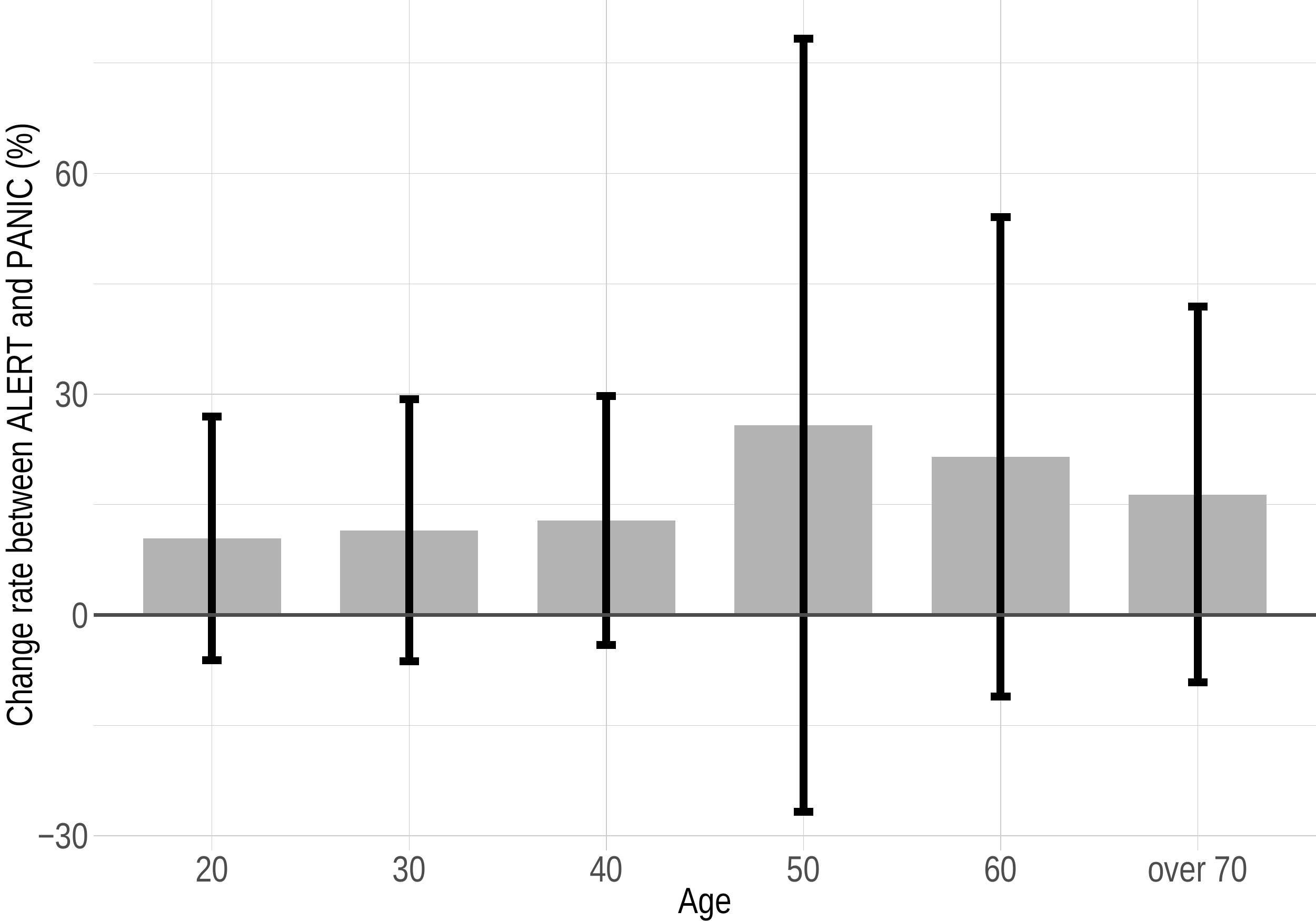}}
    \caption{The hourly mobility patterns by gender and age.}
    \label{fig:rq3}
\end{figure}

\subsection{RQ3: How are demographic characteristics related to human mobility during COVID-19?}

RQ3 is answered through analyzing the mobility patterns by gender and age during COVID-19. According to a study that assesses risk factors of COVID-19 \citep{jordan2020covid}, the age of more than 60 is the most vulnerable demographic group. Therefore, we analyze the most vulnerable age band in conjunction with gender in Figs.~\ref{fig:rq3} (a) and (b). 
We do not consider land-use and socio-economic factors for this analysis but show the mobility averaged over all cases.
Interestingly, the male population in their 60s or more shows greater mobility in general than females of the same ages.
This might be due to the cultural embeddedness of traditional gender roles in mobility patterns, where men tend to work outside while women work as housewives.
The changing gender roles in younger generations can be observed in the mobility patterns of people in their 20s, where the mobility difference between men and women is minimal (Fig.~\ref{fig:rq3} (c) and (d)). 

Another finding is that the holiday mobility of males in their 60s or more increases during PANIC and afterward. 
The mobility of females also increases but not as much as that of males.
To understand the increase of mobility in old generations better, we plot the mobility change by age group as bar charts in Figs.~\ref{fig:rq3} (e) and (f).
These figures show the average mobility changes (\%) and standard deviations as error bars in the residential areas reported in Fig.~\ref{fig:rq1} (a) and Fig.~\ref{fig:rq1-2} (a), respectively. 
Similar to the patterns observed in Fig.~\ref{fig:rq1} (a), the overall mobility of all age groups decreases during PANIC with varying degrees.
In the cells with increased mobility, while they are not statistically significant due to the small number of cells, older generations tend to move more than other age groups.
These observations indicate that the older generations show less mobility change in many areas after the COVID-19 outbreak and sometimes present even increased mobility compared to younger groups. 
This pattern is counter-intuitive to what we have reviewed in Section~\ref{sec:determinants}, where we expected that the older generations' perceived risk of COVID-19 through news articles might hamper their mobility. 
This result implies that older generations might be more vulnerable than expected not only due to their age from a medical perspective \citep{jordan2020covid}, but also because of the mobility pattern that does not show a meaningful decrease during the pandemic.

\subsection{RQ4: How did human mobility change during COVID-19 in the regions with important POIs and on the national election day?}

% Case studies on mobility change in ICN airport, massive transit stations, and other POIs. 

\begin{figure}[t]
    \centering
    \subfigure[The hourly mobility in Incheon National Airport.]{\includegraphics[width=0.49\textwidth]{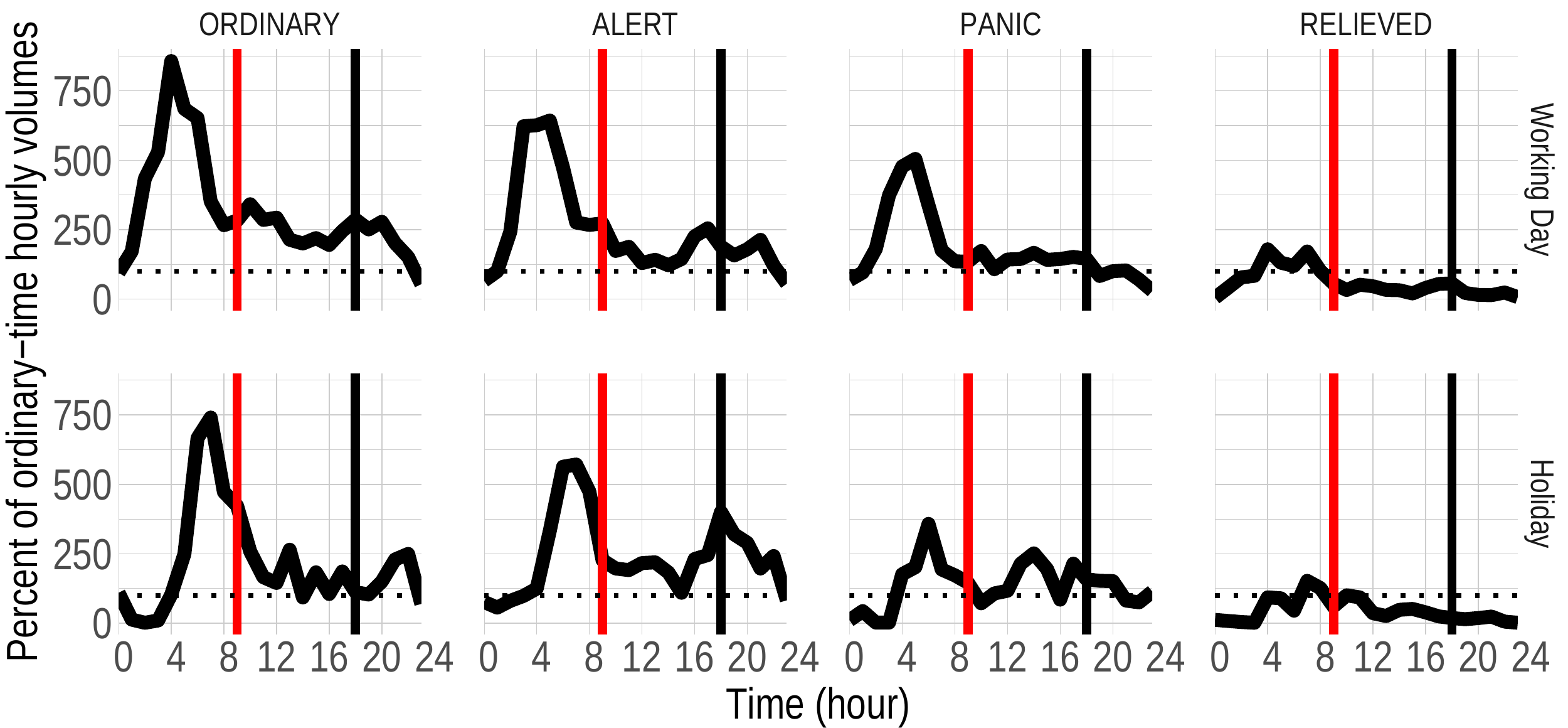}}
    \subfigure[The hourly mobility in Incheon Bus Terminal Station.]{\includegraphics[width=0.49\textwidth]{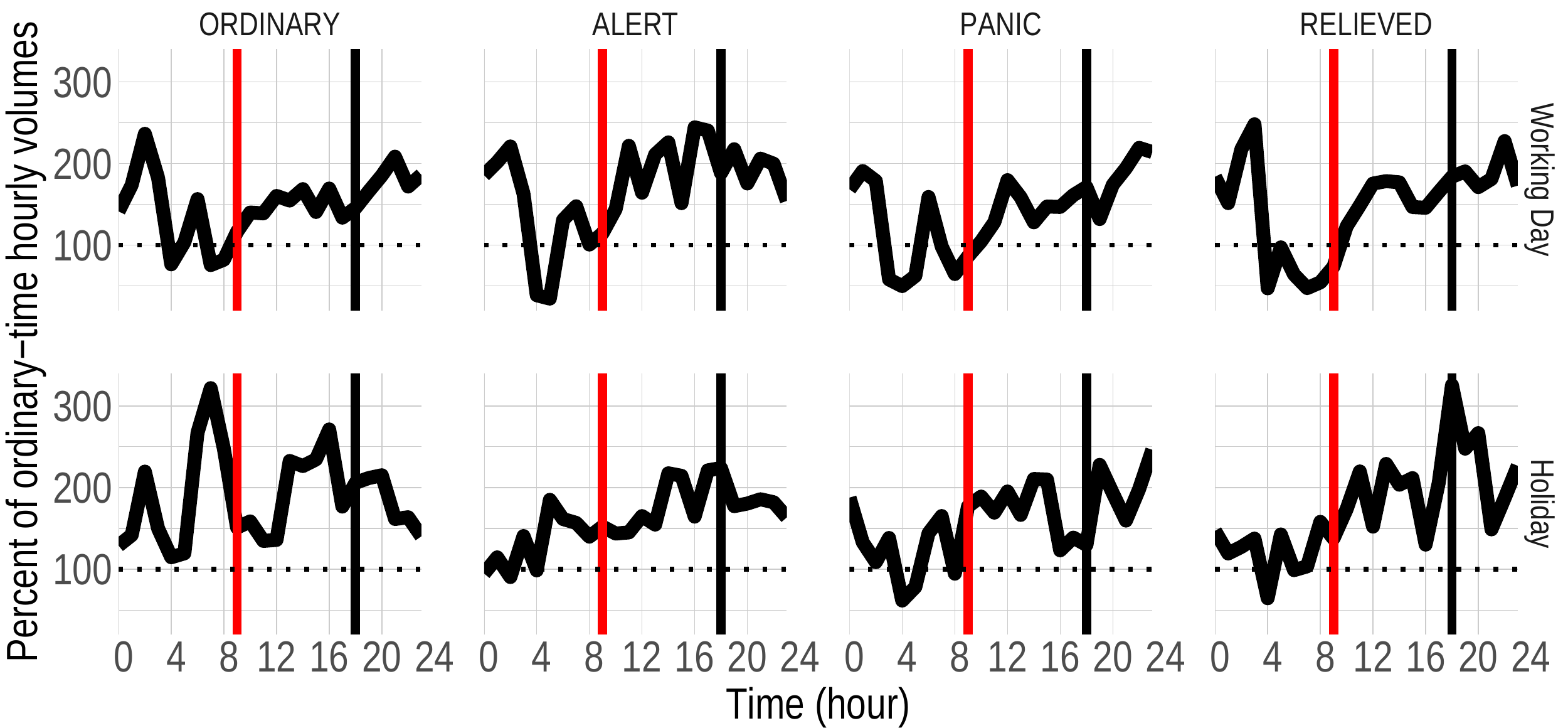}}
    \caption{The hourly mobility patterns in major transportation-related cells.}
    \label{fig:rq4-1}
\end{figure}

\paragraph{Passenger Transportation. }Fig.~\ref{fig:rq4-1} shows the hourly patterns in the two Voronoi cells related to mid- to long-distance transportation. 
The mobility around Incheon National Airport drastically decreases after the first massive infections (-31.46\%, $p$=8.24e-6). 
Compared to ORDINARY, the mobility in RELIEVED is almost four times smaller (-80.50\%, $p$=1.21e-56). This is because many international flights have canceled or temporarily stopped services; international travelers have also decided to stay in the country due to the global pandemic. 

In the cell with the city's major ground transportation, however, the mobility pattern decreases only on working days during PANIC and RELIEVED (-17.76\%, $p$=1.10e-4).\footnote{Incheon does not have high-speed train stops (whereas Seoul has three and a nearby city has one). Thus, people should either move to Seoul to take trains or use the bus terminal to travel to other cities. Therefore, we characterize this cell as the major domestic transportation method.} 
On holidays, mobility change shows two interesting patterns. 
During the afternoons and night times, the mobility increases during PANIC; conversely, during the morning and dinner times, the mobility decreases sharply. 
This might indicate that mid- to long-term travels to other cities increased while short-term trips decreased.
In RELIEVED, the overall mobility in holidays significantly increases (14.19\%, $p$=0.04) compared to ALERT and PANIC. 
The sharp decrease in morning- and dinner-time mobility observed during PANIC gets back or even increases in RELIEVED. 
This could be the case that people who left the city for mid- to long-term come back to the city, and short-term travelers re-start traveling to other cities after observing the golden-cross.

\begin{figure}[t]
    \centering
    \includegraphics[width=0.49\textwidth]{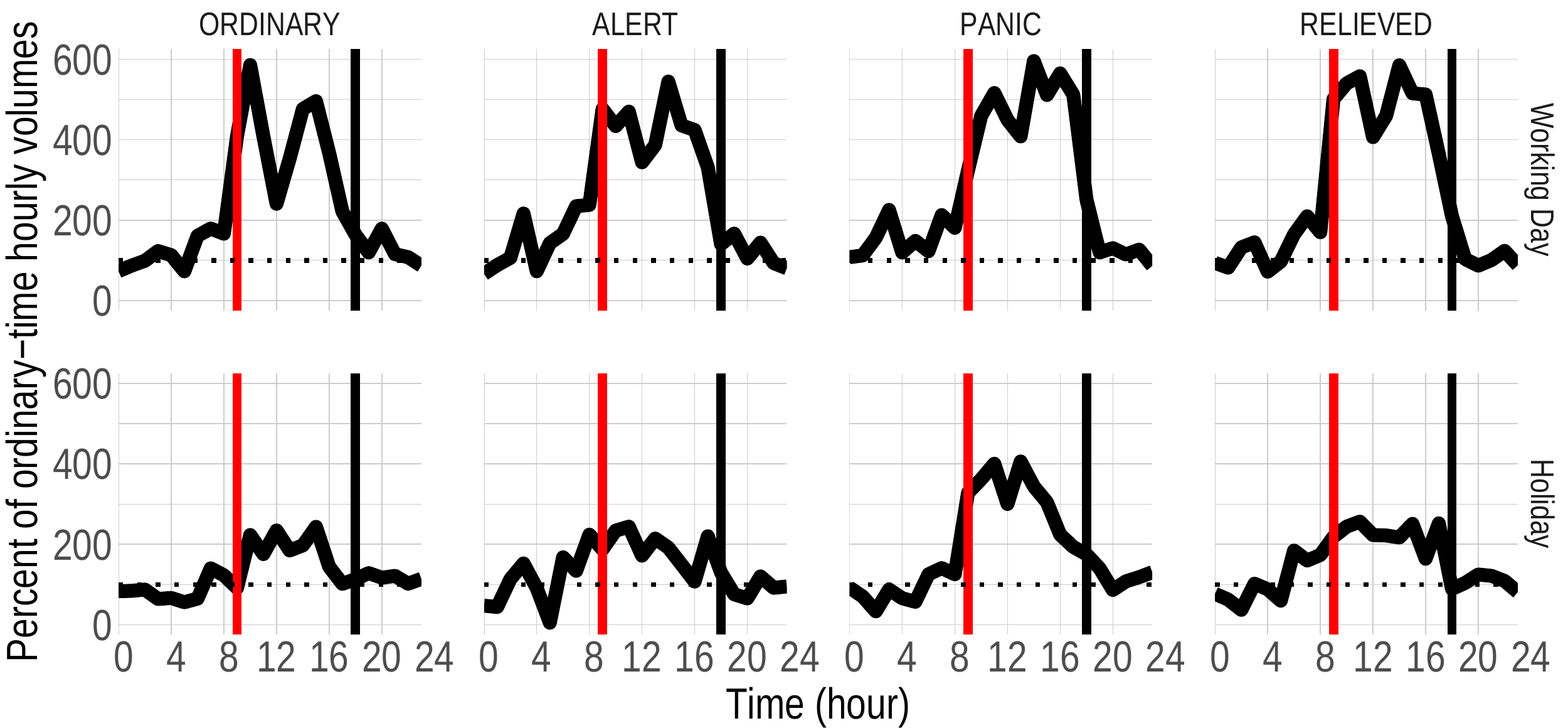}
    \caption{The hourly mobility patterns in hospitals}
    \label{fig:rq4-2}
\end{figure}

\paragraph{Healthcare Services. } There are a couple of general hospitals in the city, such as Inha University Medical Center, Catholic University Medical Center, and Gacheon University Gil Medical Center.
Fig.~\ref{fig:rq4-2} shows the mobility pattern in the cells with these hospitals. 
Because they are the largest hospitals in the city of Incheon and are responsible for taking care of COVID-19 patients, their mobility increases for all cases (24.00\%, $p$=6.67e-5). Only the holiday mobility in RELIEVED is similar to that in ALERT, which indicates a decreased number of COVID-19 patients. 
Around the golden-cross, the number of daily new patients is no more than 50 nationwide. 
Therefore, it is possible to say that mobility in hospital regions decreases as COVID-19 patients decrease after the golden-cross.

\paragraph{Education. }The city of Incheon is one of the education-oriented cities in South Korea. There are many higher education institutions, such as George Mason University Korea, The University of Utah Asia Campus, The State University of New York Korea, Yonsei University Global Campus, Inha University. 
All those universities except Inha University are located close to each other across adjacent Voronoi cells. They are not open for Spring 2020.

\begin{figure}[t]
    \centering
    \subfigure[The hourly mobility in Inha University]{\includegraphics[width=0.49\textwidth]{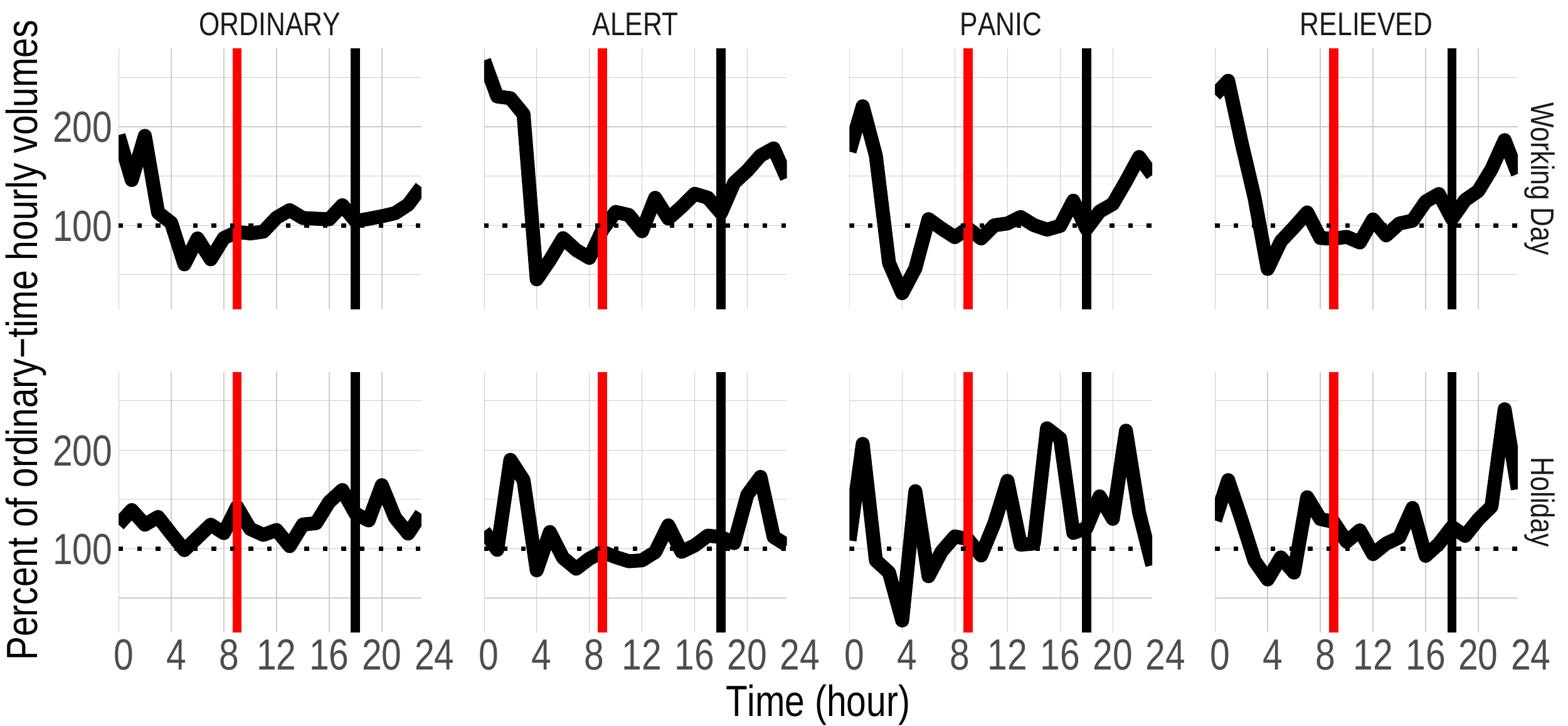}}
    \subfigure[The hourly mobility in other universities in the Songdo district]{\includegraphics[width=0.49\textwidth]{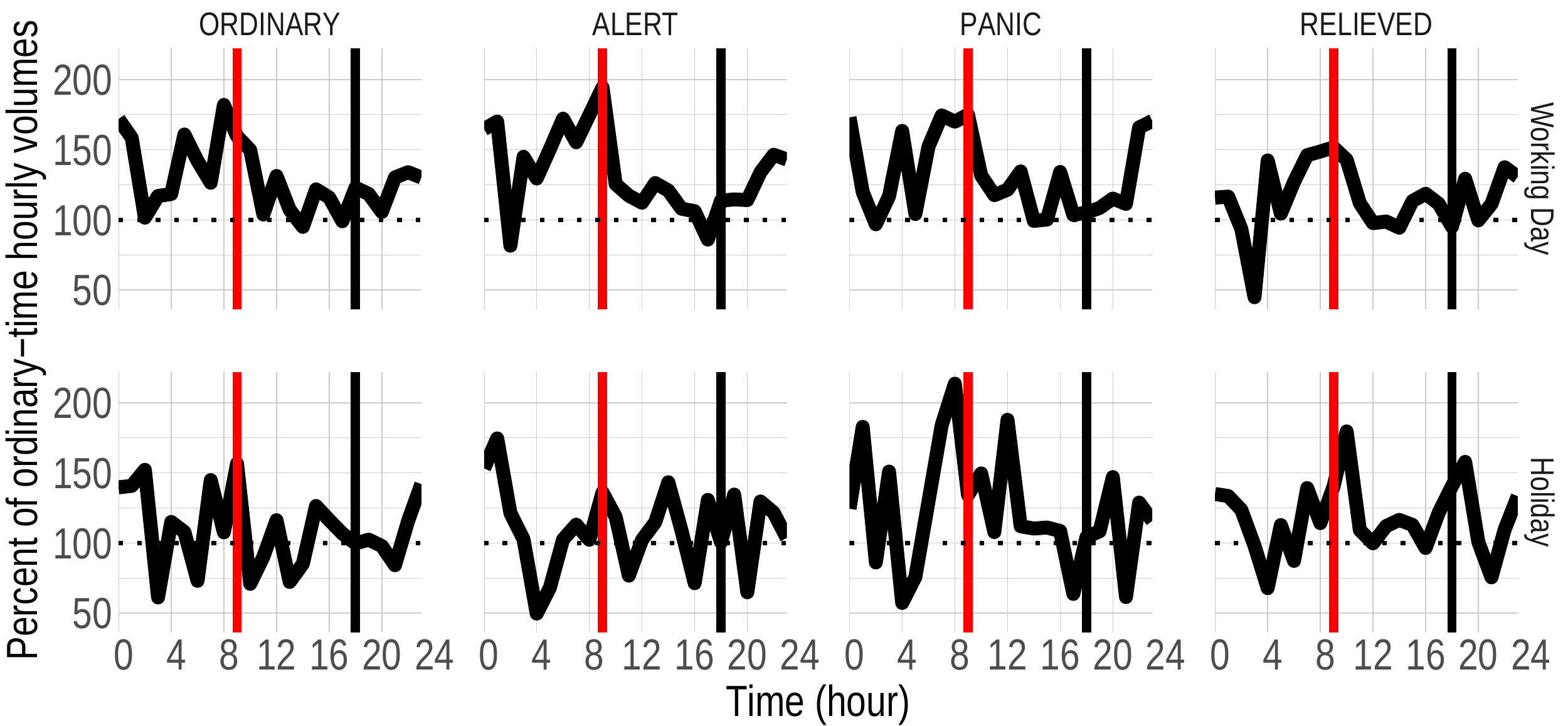}}
    \caption{The hourly mobility patterns in universities}
    \label{fig:rq4-3}
\end{figure}

Fig.~\ref{fig:rq4-3} shows that mobility decreases slightly during the working days in PANIC around Inha University (-12.07\%, $p$=0.02). 
However, its holiday mobility significantly increases during PANIC because this university opens its campus to residents (27.64\%, $p$=0.031). 
The university campus might have played a role as a park during PANIC.  

For other universities located in the Songdo district where premium, apartments, and parks are co-located, their mobility patterns do not change much on working days in PANIC. 
People visit the campuses on holiday mornings during PANIC (46.60\%, $p$=0.028). Because the Songdo district, highlighted in a blue rectangle in Fig.~\ref{fig:volume} (a), also has accessible parks, recreational zones, and commercial facilities, their working day mobility does not decrease in general.
In both figures, one interesting pattern is that the high holiday mobility in PANIC disappears in RELIEVED. This might indicate that people are getting back to normal.

\begin{figure}[ht]
    \centering
    \subfigure[Bupyeong area, the area with the largest number of voters (450K)]{\includegraphics[width=0.49\textwidth]{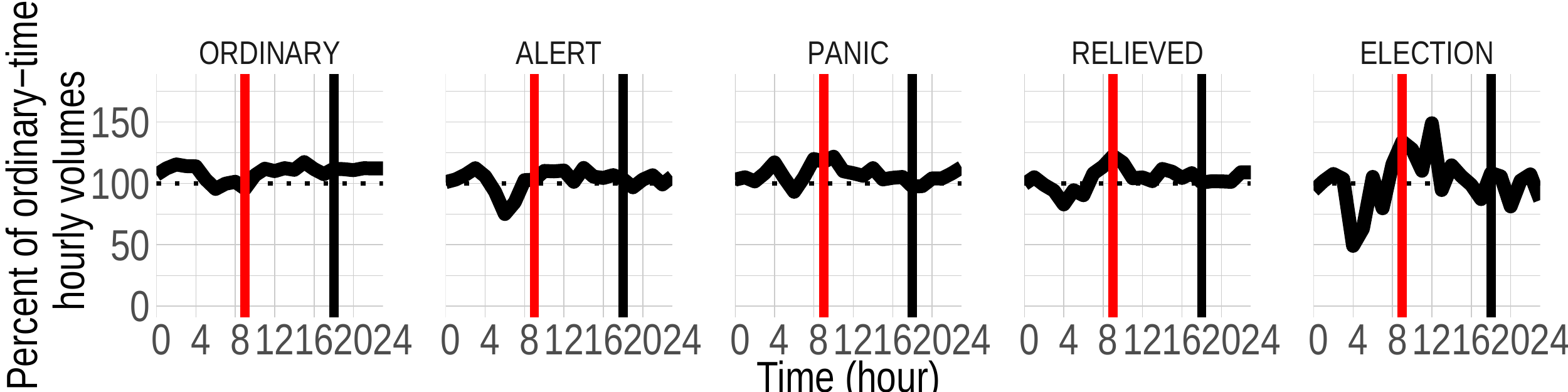}}
    \subfigure[Yeonsu area, the area with the smallest number of voters (250K)]{\includegraphics[width=0.49\textwidth]{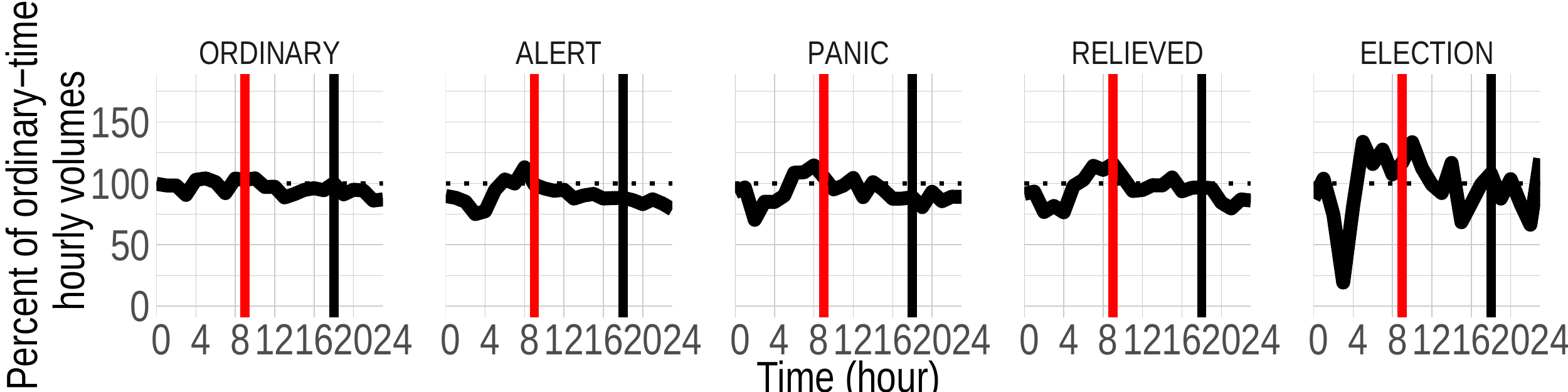}}
    \subfigure[Top-3 cells in housing price]{\includegraphics[width=0.49\textwidth]{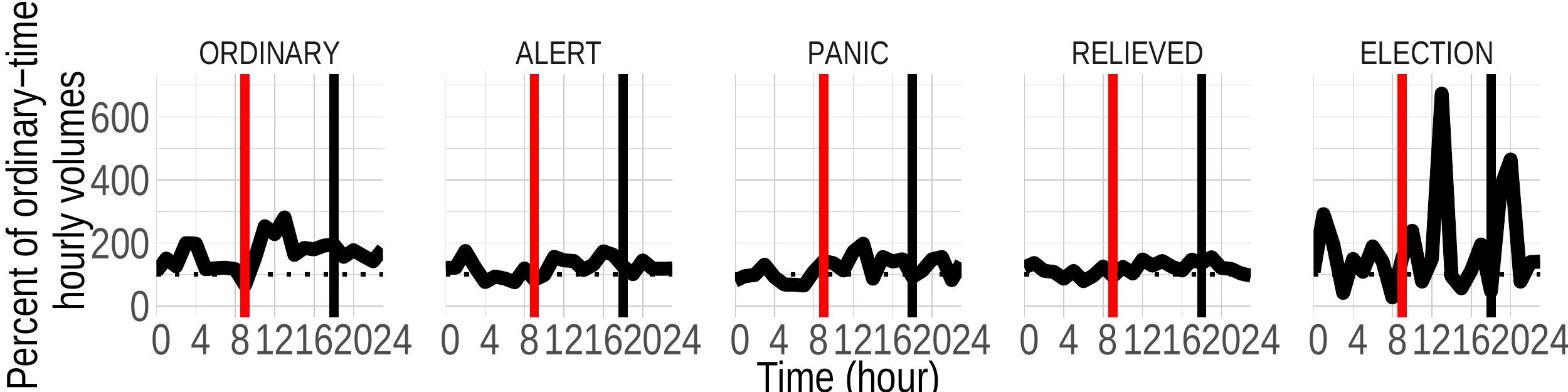}}
    \subfigure[Bottom-3 cells in housing price]{\includegraphics[width=0.49\textwidth]{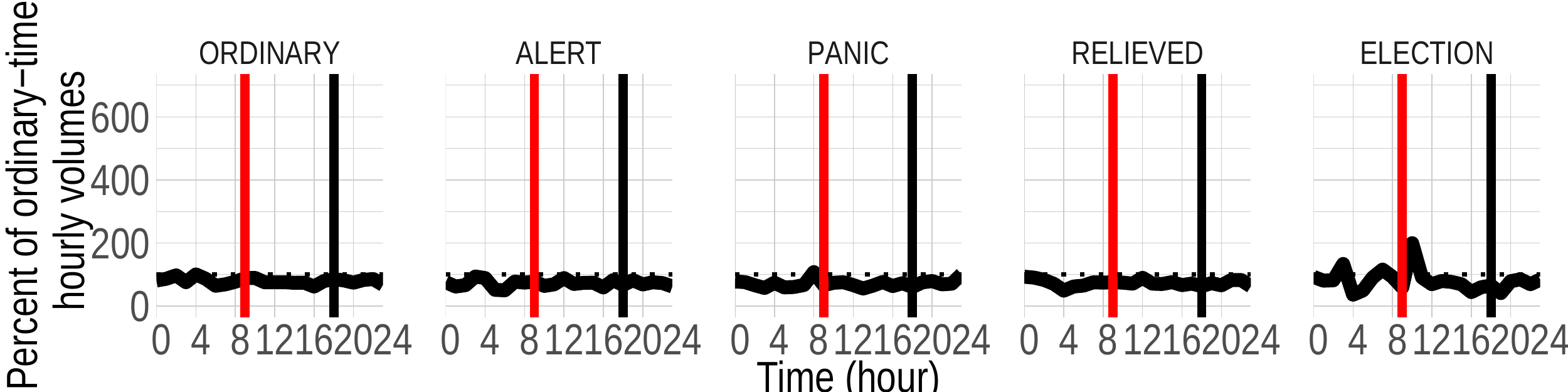}}
    \caption{The hourly mobility patterns on the legislative elections day. We compare with the holiday mobility in other periods.}
    \label{fig:rq4-4}
\end{figure}

\paragraph{Legislative  Elections. } South Korea's 21st legislative elections were held on April 15th, a temporary holiday promoting voting. They were the first nationwide event since the COVID-19 outbreak in South Korea. As shown in Fig.~\ref{fig:rq4-4}, the mobility on the election day significantly increases in all cells related to the events, compared to the mobility in other holidays. 
Fig.~\ref{fig:rq4-4} (a) shows the mobility pattern in the area with the largest number of voters, and (b) shows the smallest one. Their increase ratios are 10.18\% ($p$=0.024) in Fig.~\ref{fig:rq4-4} (a) and 7.03\% ($p$=0.045) in Fig.~\ref{fig:rq4-4} (b).

Another interesting finding on the election day is the difference in mobility by wealth. In Fig.~\ref{fig:rq4-4} (c), people in the top-3 areas (in terms of housing prices) move more actively than those in the bottom-3 areas. 
This difference might provide important implications for the relationship between socio-economic status and socio-political equity, as discussed in the next section.

%%% Maybe together, after filling all the analysis
\section{A Typology of Populations during the Pandemic}

Answering research questions is scientifically meaningful because it can show mobility pattern changes that are different from or similar to countries with different social distancing measures. 
In addition to this scientific contribution in the context of mild social distancing, the findings provide design implications for future systems in the CSCW domain. 
Based on the mobility patterns identified concerning diverse factors, we discuss potential opportunities to design technological interventions that could help flatten the curve and sustain people's daily practices and work during pandemics.
As an initial effort to provide such implications, a typology of populations is identified and developed based on their mobility patterns during COVID-19 and associated factors, which imply their lifestyles and work conditions in the context of mild social distancing.

\subsection{Crowd-avoiding Outdoor Workers}

One of the salient types of people during COVID-19 is those who still commute between their homes and workplaces while changing their commuting times, which we call ``crowd-avoiding outdoor workers.''
In the cells that present an increase in mobility (Fig.~\ref{fig:rq1-2}, mobility increased in early mornings and early afternoons on working days in both residential and commercial areas.
This may mean many people who had to go outside for their work shifted their commuting times to avoid rush hours.
Their concerns might be to ensure their safety during commuting times. 
For this type of worker, congestion- and crowd-monitoring systems or route recommendation/coordination systems could help mitigate their safety concerns during pandemics.

This difficulty is observed in cells with low socio-economic status (Fig.~\ref{fig:rq2}). 
While other patterns did not change between ALERT and PANIC, cells with low socio-economic status present peaks during morning times and in the afternoons, just like working days. 
This means many people may have to work more during the holidays in PANIC than ORDINARY, maybe because they need to complement decreased income through gigs or part-time jobs.
We hypothesize that this pattern is not due to other leisure or religious activities because they are not observed during ORDINARY.  
Further research is needed to understand the citizen groups who present this mobility pattern through surveys or in-depth interviews. 
If the reason for the increased mobility is because of extra work, the design of crowd-monitoring systems could be improved by considering the job characteristics of crowd-avoiding outdoor workers.

While not designed in this context, there have been relevant geographically-oriented recommendation systems developed in the CSCW and related fields. 
For example, Priedhorsky and Terveen developed a crowdsourcing system called Cyclopath to recommend bicycle routes for cyclists in the City of Minneapolis, Minnesota, based on people's biking trajectory data \citep{priedhorsky2008computational}.
Also, Quercia and colleagues designed happy maps that recommend aesthetically pleasant routes by using people's ratings of urban scenes \citep{quercia2014shortest}. 
Although these systems focus on geographical routes, a similar approach could be used to recommend a route and departure time to help workers safely commute to their workplaces dynamically.

\subsection{Old Workers: Poverty of the Already Vulnerable Population?}

Related to workers who have to shift their commuting times, old generations in South Korea could be another population that needs special consideration from a mobility perspective.  
One inference that could be made based on the age group observation is that a large portion of the older generations who are more vulnerable to COVID-19 than other generations might still need to commute for their work during the pandemic.
As we observed in Fig.~\ref{fig:rq2}, people who are in their 60s or more present a smaller decrease in mobility during the pandemic compared to people who are in their 20s to 50s.  
This might be because, in South Korea, people have to retire around the age of 60 by law, and many rely on a gig economy or part-time jobs after retirement.  

Studies and news articles show that these generations (usually characterized as ``baby boomers'' and ``Korean-war generations'') work hard after their retirements but still in poverty \citep{old_poverty,jang2019}. 
A news outlet reported that 33\% of the retired population in South Korea still work but remain in a low socio-economic status, which is one of the long-standing social issues \citep{old_poverty}.
A research article published by Korea Institute of Finance (KIF) reports that the proportion of the old generations in the relative poverty level is 43.8\% and that in their absolute poverty level is 32.6\%, which is a significantly higher percentage than the absolute poverty rate of younger generations at 9.3\% \citep{jang2019}.
According to this report, the reasons for the low socio-economic status include the massive retirements of people who are baby boomers, a lack of income sources for the old generations, insufficient preparations for retirement life, and insufficient retirement funds.

From a technological intervention perspective, providing a crowd-monitoring system is not sufficient for this population. 
Technological designs need to consider not only the safe commuting of this population during pandemics, but also their safe job information and potential digital divides such as technical literacy, data literacy, and physical access to the internet \citep{van2012evolution}.
At the same time, policies and social supports need to be combined for the welfare of the old generations because of their double hardships during the pandemic: both old generations' medical vulnerability and work conditions are related to their socio-economic status. 
Health authorities and policy-makers can benefit from this analysis in implementing social support policies by knowing how hardships during COVID-19 manifest across generations in people's mobility patterns.

\subsection{Working Voters}
The difference between regions with high and low socio-economic status also manifests in the mobility patterns on the legislative election day, as depicted in Fig.~\ref{fig:rq4-4}. 
In the cells with low socio-economic status, the mobility change is minimal on the election day compared to wealthy areas, and their peak in mobility is concentrated in morning times. 
This may indicate that a non-trivial portion of them voted in the morning and moved to other cells to work even though it is an official holiday, or they voted in advance.\footnote{It was possible to vote in advance a few days earlier than the official legislative election day.}
Either way, this phenomenon provides implications for the design of technological interventions for elections during pandemics. 
On the one hand, the voting, in general, can benefit from a design of technological coordination systems that help minimize the congestion at voting venues by assigning voters to appropriate time slots. 
In this case, the designers of such systems need to consider existing voting methods and known factors that could shape voting turnouts, such as early in-person voting and voters' demographic characteristics \citep{fitzgerald2005greater}.

On the other hand, the mobility-driven characteristics of voters need to be taken into accounts, such as their available times and routes on the election and pre-election days. 
As the mobility change patterns indicate, there might be many citizens who still need to work even on voting days, e.g., self-employed people.
Safety issues during a voting process can result in small voting turnouts, particularly in regions with low socio-economic status due to their limited voting time slots. Any interventions for voting during pandemics would need to be designed in a way that does not decrease the voting rate among this population. 
At the same time, policy-makers and National Election Commission could benefit from the mobility pattern data by choosing appropriate voting locations for the convenience and safety of marginalized populations.

\subsection{Flexible Office Workers}

A noticeable pattern observed from the data analysis is a decrease in mobility in the afternoons in residential, commercial, and industrial cells on working days (Fig.~\ref{fig:rq1}). 
During the pandemic, people in these cells decreased their outside activities in the afternoons and rush hours.
Because there is no peak during the rush hours around dinner time, people may not have gone to work during the pandemic. 
Based on these patterns, it is possible to infer that a large group of people are flexible in changing their working times and locations, such as opting to work from home.
For this group of workers, their primary focus might be productivity. 
If so, asynchronous, remote collaboration tools might be helpful. 

As briefly mentioned in Section~\ref{sec:intro}, technological tools for facilitating remote collaboration are among the core topics in the fields of CSCW and HCI. 
Bj{\o}rn and colleagues report, through their ethnographic work, that successful remote collaboration should involve (1) the development of common ground among team members to have a shared understanding of concepts, vocabularies, and norms, (2) collaboration readiness such as languages, necessary skill sets, and peripheral knowledge outside of their core jobs shaped by organizational structure, (3) collaboration technology readiness such as conferencing tools and shared document practices, (4) coupling of work that creates interdependencies among remote workers that facilitate communications with and awareness of each other, and (5) organizational management that tends to be more difficult in a remote environment because of the reinforced power structure driven by the physical separation of teams \citep{bjorn2014does}.

In addition to these already complex challenges in remote collaboration, the context of pandemics can make it more difficult to develop any of these components. 
For example, a sudden change in working environments could result in a lack of preparation time for organizations and workers to develop common ground among team members, equip them with the necessary technology for remote collaboration, and structure the work in a way that creates higher interdependencies among them.
Also, remote workers could struggle from additional challenges in a home environment if they have family duties, such as taking care of children and helping household work, as reported by many recent studies \citep{cluver2020parenting}.
Accordingly, the design for remote collaboration during pandemics should consider these factors as well as people's daily mobility patterns (i.e., high in the mornings and low in the afternoons) to understand their behaviors and daily challenges better.

\subsection{Leisure Time Seekers: Socializing Goers and Day Travellers}
Part of the remote work collaboration is closely related to their socializing and leisure behaviors, because knowing and understanding team members is a good way to develop common grounds. 
Beyond this, socializing and leisure activities with friends and family members is an essential part of life for people's mental/physical health and the development of trustworthiness within a community.
According to Fig.~\ref{fig:rq1}, people usually hang out with friends and family in the afternoons of holidays in ORDINARY. 
However, in the PANIC period, the peak during dinner times disappears. 
This change suggests that socializing activities outside of their homes decreased significantly during the COVID-19 outbreak. 

Similarly, in ORDINARY of Fig.~\ref{fig:rq4-2} (b), bus stations are peaked in the mornings and in the dinner times on holidays. 
This might be because of day travelers during the holidays.
Domestic travels from Incheon to locations outside of the city are easier with buses rather than with trains due to the location of the city. 
Mobility is decreased during morning and dinner times on the holidays because day travelers who regularly go to hiking and leisure activities outside of the city might have stopped having trips during the pandemic.  
Since socializing and leisure activities are closely related to people's mental and physical health, these mobility patterns suggest a need for technological interventions to help them socialize with each other and enjoy their outdoor activities while ensuring safety.
One possible implication might be, similar to that for crowd-avoiding outdoor workers, a consideration of crowdedness in POIs or leisure places: systems could be designed by focusing on safe alternatives and times for outdoor activities.
Also, existing socializing and online networking apps could be re-appropriated by considering their emotional changes and safe times for connecting socializing goers during dinner times on holidays.

\section{Limitations and Conclusion}

As with other data-driven studies, this analysis has some limitations. 
Even though spatial and temporal units were aggregated to some degree, there is still uncertainty embedded in the processed data due to the uneven distribution of taxis across times and locations. 
Refining aggregation techniques and other imputation methods could help minimize the spatio-temporal bias. 
Another difficulty is that taxis cannot capture mobility if some regions are not accessible by cars. 
We overcame this by setting the size of Voronoi cells big enough in such areas.
However, future studies may need to take this uncertainty into account to identify mobility changes more precisely. 
Also, mixed land-use needs to be considered in future analysis. 
In this paper, we used the highest-ranking types among the manually-coded land-uses.
However, this method still has a limitation from an analytical perspective.
Developing ways to consider the mixed land-use types would make this kind of study more accurate in interpreting results, especially in cities with a high population density.

Statistical models could be improved in future studies.
This study provides a descriptive analysis of mobility changes, rather than fitting novel statistical models such as Event Study analysis or hazard models because this study focuses on providing implications for technological interventions during pandemics. 
Analyzing mobility changes with advanced statistical models and spatio-temporal analysis will benefit social scientists and policy-makers with more accurate measures and monitoring technologies.  
Finally, the use of housing transaction data as a proxy of socio-economic status needs more validation. 
Although there is strong evidence that real-estate ownership is a symbolic and material wealth indicator, there might be some discrepancies between housing prices and actual wealth at the mesoscopic region level. 
Measuring and predicting socio-economic status could benefit from machine learning models and large-scale surveys in future studies. 
Also, open data support from the statistical department in the government of South Korea will help alleviate this issue. 

Nevertheless, this study provides extensive analyses of human mobility during the COVID-19 outbreak, by possibly covering the most pre- and after-pandemic periods in South Korea.
By analyzing human mobility in diverse regions and times of the day, this paper contributes to the scientific findings on mobility during COVID-19 in the context of mild social distancing. 
Also, through the examinations of daily pattern changes in mobility, this paper identifies a typology of five mobility groups who potentially need more attention for their safe navigation of daily practices. 
We hope that the typology of populations and their discussions provide useful implications for the design of technological interventions to help relieve the on-going challenges and plan for the post-pandemic era.

% \begin{acks}
% To be added.
% \end{acks}

\bibliographystyle{ACM-Reference-Format}
\bibliography{sample-acmsmall}

\end{document}